\documentclass[12pt,draftclsnofoot,onecolumn]{IEEEtran}
\ifCLASSINFOpdf
\else
\fi
\usepackage{amsmath,dsfont,bbm,epsfig,amssymb,amsfonts,amstext,verbatim,amsopn,cite,subfigure,multirow,multicol,lipsum,xfrac}
\usepackage{amsthm}
\usepackage{mathtools,amsthm}
\usepackage{perpage}
\usepackage{balance}
\usepackage{url}
\usepackage{amsfonts}
\usepackage{epsfig}
\usepackage[font={small}]{caption}
\usepackage{psfrag}	
\usepackage{etoolbox}
\usepackage{algorithmicx}
\usepackage[Algorithm,ruled]{algorithm}
\usepackage{algpseudocode}
\usepackage{pifont}
\usepackage[utf8]{inputenc}
\usepackage[T1]{fontenc}  
\usepackage[nolist]{acronym}
\usepackage{paralist}
\usepackage{enumitem}
\usepackage{bbm}
\usepackage[process=auto]{pstool}
\usepackage{tikz}
\usetikzlibrary{shapes,arrows}
\newcommand{\setX}{\mathbbmss{X}}

\newcommand{\setR}{\mathbbmss{R}}

\newcommand{\setS}{\mathbbmss{S}}
\newcommand{\setW}{\mathbbmss{W}}

\newcommand{\setZ}{\mathbbmss{Z}}

\newcommand{\setC}{\mathbbmss{C}}

\newcommand{\setT}{\mathbbmss{T}}

\DeclareMathOperator*{\argmin}{argmin}

\newcommand{\av}{\mathrm{Erg}}
\newcommand{\snr}{\mathrm{SNR}}

\newcommand{\rmp}{\mathrm{p}}

\newcommand{\rmF}{\mathrm{F}}

\newcommand{\rmR}{\mathrm{R}}

\newcommand{\rmG}{\mathrm{G}}
\newcommand{\rmQ}{\mathrm{Q}}

\newcommand{\rmI}{\mathrm{I}}
\newcommand{\rmh}{\mathrm{h}}

\newcommand{\rmj}{\mathrm{j}}

\newcommand{\her}{\mathsf{H}}
\newcommand{\glse}{\mathrm{glse}}
\newcommand{\dec}{\mathsf{dec}}

\newcommand{\sfM}{\mathsf{M}}

\newcommand{\sfD}{\mathcal{D}}

\newcommand{\sfR}{\mathsf{R}}
\newcommand{\sfd}{\mathsf{d}}

\newcommand{\sfp}{\mathsf{p}}
\newcommand{\sfc}{\mathsf{c}}
\newcommand{\sfa}{\eta}

\newcommand{\mar}{\mathcal{R}}
\newcommand{\maf}{\mathcal{F}}

\newcommand{\mae}{\mathcal{E}}
\newcommand{\maz}{\mathcal{Z}}
\newcommand{\mas}{\mathcal{S}}

\newcommand{\mam}{\mathcal{M}}
\newcommand{\mad}{\mathcal{D}}

\newcommand{\mg}{\mathcal{G}}
\newcommand{\mai}{\mathcal{I}}

\newcommand{\bvv}{\mathbf{v}}

\newcommand{\mh}{\mathbf{h}}

\newcommand{\bx}{{\boldsymbol{x}}}

\newcommand{\vv}{\mathrm{v}}

\newcommand{\xx}{\mathrm{x}}

\newcommand{\set}[1]{\left\lbrace#1\right\rbrace}
\newcommand{\prant}[1]{\left( #1 \right)}
\newcommand{\cnor}[1]{\mathcal{CN}\left( \boldsymbol{0}, #1 \right)}

\newcommand{\bz}{{\boldsymbol{z}}}

\newcommand{\bs}{{\boldsymbol{s}}}

\newcommand{\bv}{{\boldsymbol{v}}}

\newcommand{\dif}{\mathrm{d}}

\newcommand{\by}{{\boldsymbol{y}}}

\newcommand{\trp}{\mathsf{T}}

\newcommand{\mA}{\mathbf{A}}

\newcommand{\mI}{\mathbf{I}}
\newcommand{\mone}{\mathbf{1}}
\newcommand{\mJ}{\mathbf{J}}
\newcommand{\mG}{\mathbf{G}}
\newcommand{\mQ}{\mathbf{Q}}
\newcommand{\mS}{\mathbf{S}}
\newcommand{\mU}{\mathbf{U}}
\newcommand{\mD}{\mathbf{D}}
\newcommand{\mM}{\mathbf{M}}

\newcommand{\mT}{\mathbf{T}}
\newcommand{\mH}{\mathbf{H}}

\newcommand{\md}{\mathrm{D}}
\newcommand{\E}{\mathbb{E}\hspace{.5mm}}

\newcommand{\rs}{{\mathsf{rs}}}
\newcommand{\rsb}{{\mathsf{rsb}}}
\newcommand{\papr}{{\mathrm{PAPR}}}

\newcommand{\norm}[1]{\lVert #1 \rVert}
\newcommand{\re}[1]{\mathsf{Re}\left\lbrace #1 \right\rbrace}
\newcommand{\img}[1]{\mathsf{Im}\left\lbrace #1 \right\rbrace}

\newcommand{\abs}[1]{\lvert #1 \rvert}
\newcommand{\tr}[1]{\mathrm{Tr} \{ #1 \}}

\newtheoremstyle{mystyle}
  {}
  {}
  {}
  {}
  {\bfseries}
  {:}
  { }
  {}

\theoremstyle{mystyle}

\newtheorem{definition}{Definition}
\newtheorem{proposition}{Proposition}
\newtheorem{remark}{Remark}
\newtheorem{lemma}{Lemma}
\newtheorem*{prf}{Proof}
\newcounter{bar}

\begin{document}

\begin{acronym}
\acro{mimo}[MIMO]{Multiple-Input Multiple-Output}
\acro{csi}[CSI]{Channel State Information}
\acro{awgn}[AWGN]{Additive White Gaussian Noise}
\acro{iid}[i.i.d.]{independent and identically distributed}
\acro{cdf}[CDF]{Cumulative Distribution Function}
\acro{pmf}[PMF]{Probability Mass Function}
\acro{pdf}[PDF]{Probability Density Function}
\acro{ut}[UT]{User Terminal}
\acro{bs}[BS]{Base Station}
\acro{tas}[TAS]{Transmit Antenna Selection}
\acro{glse}[GLSE]{Generalized Least Square Error}
\acro{lse}[LSE]{Least Square Error}
\acro{mse}[MSE]{Mean Squared Error}
\acro{rls}[RLS]{Regularized Least Squares}
\acro{rhs}[r.h.s.]{right hand side}
\acro{lhs}[l.h.s.]{left hand side}
\acro{wrt}[w.r.t.]{with respect to}
\acro{rs}[RS]{Replica Symmetry}
\acro{rsb}[RSB]{Replica Symmetry Breaking}
\acro{papr}[PAPR]{Peak-to-Average Power Ratio}
\acro{rzf}[RZF]{Regularized Zero Forcing}
\acro{snr}[SNR]{Signal-to-Noise Ratio}
\acro{rf}[RF]{Radio Frequency}
\acro{mrt}[MRT]{Maximum Ratio Transmission}
\acro{psk}[PSK]{Phase Shift Keying}
\acro{lmsrf}[LMSRF]{Load-Modulated Single RF}
\acro{ofdm}[OFDM]{Orthogonal Frequency Division Multiplexing}
\end{acronym}

%
\title{GLSE Precoders for Massive MIMO Systems: Analysis and Applications}

\author{
\IEEEauthorblockN{
Ali Bereyhi,
Mohammad Ali Sedaghat,
Ralf R. M\"uller,
Georg Fischer
}
\thanks{Parts of this work have been presented in 21th International ITG Workshop on Smart Antennas (WSA) and 2017 IEEE Inter- national Symposium on Information Theory (ISIT) \cite{bereyhi2017nonlinear,bereyhi2017asymptotics}.}
\thanks{This is the extended version of the paper submitted to IEEE Transactions on Wireless Communications.}
\thanks{Ali Bereyhi and Ralf R. M\"uller are with the Institute for Digital Communications (IDC),~Friedrich-Alexander Universit\"at Erlangen-N\"urnberg (FAU), Erlangen, Bavaria, Germany (e-mails: ali.bereyhi @fau.de, ralf.r.mueller@fau.de).}
\thanks{Mohammad Ali Sedaghat is with the Cisco Optical GmbH, N\"urnberg, Bavaria, Germany (e-mail: msedaghat@cisco.com). During this work, he was with IDC, FAU, Erlangen Germany.}
 \thanks{Georg Fischer is with the Lehrstuhl f\"ur Technische Elektronik (LTE), Friedrich-Alexander Universit\"at Erlangen-N\"urnberg (FAU),  Erlangen, Bavaria, Germany (e-mail: georg.fischer@fau.de).}
\thanks{This work was supported by Deutsche Forschungsgemeinschaft (DFG) under Grant No. MU 3735/2-1.}
}


%


\IEEEoverridecommandlockouts

\maketitle

\begin{abstract}
This paper proposes the class of Generalized Least-Square-Error (GLSE) precoders for multiuser massive MIMO systems. For a generic transmit constellation, GLSE precoders minimize the interference at user terminals assuring that given constraints on the transmit signals are satisfied. The general form of these precoders enables us to impose multiple restrictions at the transmit signal such as limited peak power and restricted number of active transmit antennas. Invoking the replica method from statistical mechanics, we study the performance of GLSE precoders in the large-system limit. We show that the output symbols of these precoders are identically distributed and their statistics are described with an equivalent scalar GLSE precoder. Using the asymptotic results, we further address some applications of the GLSE precoders; namely forming transmit signals over a restricted alphabet and transmit antenna selection. Our investigations demonstrate that a computationally efficient GLSE precoder requires $41\%$ fewer active transmit antennas than conventional selection protocols in order to achieve a desired level of input-output distortion.
%
%
%
\end{abstract}
\IEEEpeerreviewmaketitle

\section{Introduction}
In Massive \ac{mimo} systems \cite{marzetta2010noncooperative,hoydis2013massive,larsson2014massive,rusek2013scaling}, performance gains such as spectral efficiency, rate reliability, and energy efficiency are enhanced at the cost of employing a large number of antennas. This rises several challenges in practice which are roughly divided into \textit{physical} and \textit{operational} ones. Physical issues, e.g., having a large antenna array within a rather small physical platform, are effectively addressed by shifting to the millimeter wave spectrum \cite{rappaport2013millimeter}. Operational bottlenecks however have not been yet precisely studied in this context. An example is the high overall \ac{rf}-cost of massive \ac{mimo} systems which makes them costly to implement. In contrast to physical challenges, operational bottlenecks could be overcome by effective design of inexpensive system modules and employing more advanced algorithms to compensate for their limited abilities~\cite{di2014spatial,liang2014low,sohrabi2016hybrid,muller2014load,sedaghat2014novel}.
%

One of the key modules in \ac{mimo} systems is the precoder which maps the data signal to a precoded signal in order to compensate the distortion caused by the channel. In fact, due to the restricted computational capacity and limited power supply at user terminals, these systems are wanted to shift most of the processing load towards the transmit side. Conventional approach for precoding are to either utilize simple linear precoders, e.g., \ac{mrt} and \ac{rzf} \cite{peel2005vector}, with low computational complexity, or to employ more advanced nonlinear schemes such as Tomlinson-Harashima \cite{fischer2002space} and vector precoding \cite{peel2005vector} at the expense of higher complexity. 

Considering either of these approaches, the main restriction on the output signal is to have limited average transmit power, and other available operational challenges are not addressed. There are therefore different bodies of work in the literature which try to address more restrictions at the precoding stage. An example is \cite{mohammed2013per} where the authors introduced the nonlinear ``per-antenna constant envelope precoding'' to construct transmit signals with constant amplitudes, i.e., unit \ac{papr}. Such approaches mainly focus on particular limitations and try to adopt prior schemes to a broader set of signal constraints. In this manuscript, we intend to develop a classical framework which addresses a general operational restriction.

\subsection{Precoding via Regularized Least Squares Regression}
The key contribution of this study is to deviate from the conventional approach and consider precoding as a constrained optimization problem. In this respect, precoding is interpreted as a signal shaping problem for which several classical approaches can be employed. 

In order to illustrate this latter point, consider a basic multiuser downlink \ac{mimo} setting with $K$ single-antenna users and $N$ transmit antennas at the base station. Let $\mH\in\setC^{K\times N}$ represent the channel coefficients. For this setting, a precoding scheme is a transform which maps user data symbols, collected in $\bs=[s_1, \ldots, s_K]^\trp \in \setC^{K \times 1}$, into an $N$-dimensional signal $\bx\in\setX^{N\times 1}$ where $\setX$ is the ``precoding support'' enclosing all possible transmit constellation points, e.g., for quadrature \ac{psk} transmission $\setX=\set{ \pm 1 \pm \rmj }$. In this case, the vector of the received signals at user terminals, i.e., $\by=[y_1, \ldots, y_K]^\trp\in\setC^{K\times 1}$, reads
\begin{align}
\by=\mH \hspace*{.5mm} \bx + \bz \label{eq:ch}
\end{align}
where $y_k$ denotes the symbol received by user $k$ and $\bz\in\setC^{K\times 1}$ is thermal noise.

The main objective of the precoder is to design the mapping such that the receive terminals only deal with thermal noise. In other words, the precoder aims to choose the mapping $\bs\to\bx$ which minimizes the distortion between $\mH \bx$ and $\bs$ while satisfying some given constraints on the transmit signal $\bx$. Assuming that the distortion is measured by the \ac{mse}\footnote{Note that, in general, one can consider other distortion measures, as well.}, i.e., $\norm{\mH\bx-\bs}^2$, precoding can be interpreted as a \ac{rls} problem in which, given a set of constraints, the squared error is desired to be minimized. This connection between \ac{rls} and precoding was initially observed in \cite{sedaghat2017least} where the ``nonlinear \ac{lse} precoder'' was developed to address transmit power restriction for a large class of signal constellations. For given $\bs$ and $\mH$, the nonlinear LSE precoder finds the transmit~signal~as
\begin{align}
\bx =\argmin_{\bv \in {\setX^N}} \norm{\mH \bv- \bs}^2 + \lambda \norm{\bv}^2. \label{eq:lse0}
\end{align}
The precoding scheme in \eqref{eq:lse0} can be considered as \ac{rls} regression with quadratic regularization term. In fact, this approach simultaneously minimizes the \ac{mse} term and controls the transmit power by penalizing. The main advantage of this scheme compared to conventional algorithms is the generality of the precoding support which leads to address a broader set of operational restrictions, such as transmission with limited \ac{papr} and~discrete~constellations. A similar approach has been taken for precoding with 1-bit digital-to-analog converters in \cite{jacobsson2017quantized}. 

Considering the analogy between precoding and \ac{rls} signal shaping, the \ac{rls}-based approach can further be developed to address other operational restrictions at the precoding stage. In fact, one could modify both the penalty and the precoding support in \eqref{eq:lse0} such that the signal satisfies the desired constraints. An example is limitation in the number of active transmit antennas required in some massive \ac{mimo} settings to reduce the overall \ac{rf}-cost \cite{li2014energy,asaad2017tas,tervo2017energy}. We further illustrate our precoding approach by discussing this particular example.

\subsection{Applications to Antenna Selection}
\label{sec:AntSelect}
A classical solution to high implementational \ac{rf}-costs in massive \ac{mimo} systems is antenna selection. In downlink, this task is done prior to precoding via a selection algorithm. The transmit signal is then constructed over the selected antennas.  In addition to \ac{rf}-cost reduction, antenna selection is shown to enhance the performance in some \ac{mimo} settings \cite{asaad2017optimal,xu2013energy,tervo2017energy}. In general, the optimal selection algorithm is defined based on the performance measure, e.g., the achievable sum-rate, energy efficiency or the \ac{mse}. For most metrics which quantify the performance, the optimal approach confronts us with an exhaustive search whose complexity grows exponentially with the number of transmit antennas. As the result, the optimal algorithms are infeasible for implementation and suboptimal greedy solutions are employed in practice. These alternative approaches pose polynomial computational complexity at the expense of performance degradation; see \cite{gharavi2004fast,gorokhov2003receive,bereyhi2018stepwise} for some particular greedy selection algorithms.

In general, antenna selection algorithms in a massive \ac{mimo} downlink setting, whose number of transmit antennas grows large, can be classified as follows:
\begin{enumerate}
\item The number of active antennas remains \textit{fixed} while the total number of antennas growing large, e.g., the settings considered in  \cite{li2014energy,asaad2017tas}, and
\item The number of selected antennas grows large with the total number of antennas such that the fraction of transmit antennas being active is kept fixed.
\end{enumerate}

In the former regime, the crucial question is to see whether the large-system properties, such as channel hardening, are preserved; see for example the study in \cite{asaad2017tas}. Such a question arises here, since in this regime, the number of active transmit antennas does not scale with the system dimensions, and the growth in the total number of antennas impacts the performance \textit{implicitly} through the antenna selection procedure. The investigations have demonstrated that even simple selection approaches in this regime maintain the large-system properties.

In the second regime of antenna selection, preservation of large-system properties is obvious due to the fact that the number of selected antennas grows large linearly with the total number of antennas. Therefore, the fundamental analytic task in this case is to characterize the performance of optimal antenna selection and investigate the possible degradation caused by using suboptimal selection algorithms. In contrast to the first regime, there are several questions left unanswered in this case, e.g., optimal performance for antenna selection. Such a study could help describing quantitatively the effectiveness of the state of the art algorithms.

Invoking the illustrated \ac{rls} framework for precoding, the restriction on the number of active transmit antennas can be formulated as \ac{rls} signal shaping with a \textit{sparsity} constraint. In fact, restricting the number of active transmit antennas mathematically means that a certain fraction of entries in the transmit signal $\bx$, which corresponds to passive antennas, should  be zero. From the literature of compressive sensing \cite{donoho2006compressed,candes2006robust,foucart2013mathematical}, this signal shaping problem is addressed via \ac{rls} regression with $\ell_0$-norm\footnote{By $\ell_0$-norm, we mean the number of non-zero entries in the vector.} or $\ell_1$-norm regularization. Consequently, the number of active transmit antennas can be constrained via \eqref{eq:lse0} with $\ell_0$-norm or $\ell_1$-norm penalty. 

\subsection*{Contributions and Analytic Tools}
This paper introduces a generic precoding scheme based on \ac{rls} regression. The scheme is referred to as \ac{glse} precoding and addresses a large set of instantaneous constraints on the transmit signal with antenna selection being just one of them. The \ac{glse} scheme extends the quadratic regularization in \eqref{eq:lse0} into a general penalty which encloses several transmit restrictions. As the result, the nonlinear \ac{lse} precoding in \cite{sedaghat2017least} can be considered as a special case of \ac{glse} scheme which only constrains the instantaneous transmit power. In the large-system limit, we evaluate the performance of \ac{glse} precoding by determining a lower bound on the achievable ergodic rate per user. Our analyses demonstrate that the output symbols of \ac{glse} precoders are identically distributed and follow the distribution of an equivalent scalar precoder. We refer to this property as ``asymptotic marginal decoupling property''. Using large-system results, we investigate the performance of some particular forms of \ac{glse} precoding which address \ac{tas}, \ac{papr}-limited transmission and precoding over discrete constellations. The large-system results in this paper are derived via the replica method developed in statistical mechanics. This method has been widely employed in multiuser communications, signal processing and information theory \cite{tanaka2002statistical,guo2005randomly,rangan2012asymptotic,bereyhi2016statistical,vehkapera2014analysis}. An introduction to the replica method is given in Section \ref{sec:large}.
\subsection*{Notations}
We represent scalars, vectors and matrices with non-bold, bold lower case and bold upper case letters, respectively. A $K \times K$ identity matrix is shown by $\mI_K$, and $\mH^{\her}$ indicates the Hermitian of the matrix $\mH$. The notation $\otimes$ denoted the Kronecker product. The set of real and integer numbers are denoted by $\setR$ and $\setZ$, respectively, and their corresponding non-negative subsets are shown by superscript $\left(\cdot\right)^+$; moreover, $\setC$ represents the complex plane. For $s\in\setC$, the real part, imaginary part, magnitude and phase are shown by $\re{s}$, $\img{s}$, $\abs{s}$ and $\sphericalangle s$, respectively. $\norm{\bx}$ and $\norm{\bx}_1$ denote the Euclidean and $\ell_1$-norm of the vector $\bx$, and $\norm{\bx}_0$ represents the $\ell_0$-norm defined as the number of non-zero entries. For a random variable, $\mathrm{p}(\cdot)$, $f(\cdot)$ and $\mathrm{F}(\cdot)$ represent the \ac{pmf}, \ac{pdf} and \ac{cdf}, respectively. Moreover, the expectation operator is denoted by $\E$. For brevity, the set of integers $\set{1, \ldots , N}$ is abbreviated as $[N]$ and $\phi\prant{\cdot;\rho}$ represents the \ac{pdf} of a zero-mean complex Gaussian random variable with variance $\rho$. Whenever needed, we consider the precoding support $\setX$ to be a set of discrete constellation points. Our results, however, are in full generality and hold for continuous transmit constellations as well.

\section{Problem Formulation}
\label{sec:sys}
Consider the multiuser downlink \ac{mimo} scenario given in \eqref{eq:ch} 
in which the channel matrix $\mH$, the transmit signal $\bx$ and the noise vector $\bz$ fulfill the following conditions:
\begin{enumerate}[label=(\alph*)]
\item $\mH\in \setC^{K\times N}$ represents a frequency-flat fading channel and is modeled as a random matrix whose Gramian $\mJ=\mH^{\her} \mH$ has the eigendecomposition
\begin{align}
\mJ = \mU \mD \mU^{\her}.
\end{align}
Here, $\mU_{N \times N}$ is a random unitary matrix with Haar distribution\footnote{In a nutshell, a Haar-distributed random matrix is a unitary matrix which is uniformly distributed over the unitary group. See \cite{mehta2004random} for the exact definition.}, and $\mD_{N \times N}$ is a diagonal matrix with non-zero elements $\set{\lambda_n}$ for $n\in [N]$ which denote the squared singular values of $\mH$. We assume that the empirical \ac{cdf} of the squared singular values, defined as
\begin{align}
\rmF_\mJ^{N} (\lambda)= \frac{1}{N}\sum_{n=1}^N \mone \set{\lambda_n < \lambda},
\end{align}
asymptotically converges to the deterministic \ac{cdf} $\rmF_\mJ$, i.e., $\rmF_\mJ (\lambda)=\lim\limits_{N\uparrow\infty} \rmF_\mJ^{N} (\lambda)$.

The~assumed ensemble of channel matrices encloses a large class of \ac{mimo} channel models including the well-known \ac{iid} flat Rayleigh fading and uncorrelated low-rank fading models \cite{ertel1998overview}.
\item The dimensions of $\mH$ grow large, such that the~load defined as $\alpha\coloneqq K/N$, is kept fixed in both $K$ and $N$, and is bounded from above meaning that $\alpha < \infty$.
\item The entries of $\bx_{N \times 1}$ are drawn from the support $\setX\subseteq\setC$.
\end{enumerate}
The transmit signal $\bx$ is constructed from the data symbols of the users $\set{s_1, \ldots, s_K}$ and \ac{csi} via a nonlinear \ac{glse} precoder which is defined as follows: %
\begin{definition}[\ac{glse} precoders]
\label{def:glse}
Consider a given power control factor $\rho$ and the channel matrix $\mH$. The \ac{glse} precoder with the penalty function $u(\cdot):\setX^N \mapsto \setR$ and support $\setX$ is defined~as
\begin{align}
\glse\prant{\bs|\rho,\mH}=\argmin_{\bv \in {\setX^N}} \norm{\mH \bv-\sqrt{\rho} \hspace*{.3mm} \bs}^2 +u(\bv). \label{eq:1}
\end{align}
which maps the data vector $\bs_{K\times 1}=[s_1, \ldots,s_K]^\trp\in \setC^{K\times 1}$ onto an $N$-dimensional vector. 
\end{definition}

The transmit vector, i.e. $\bx=\glse\prant{\bs|\rho,\mH}$, is given to the \ac{rf} front-end for transmission over the antennas.  We further assume that
\begin{enumerate}[label=(\alph*)]\addtocounter{enumi}{3}
\item $\bs_{K \times 1}=[s_1, \ldots,s_K]^\trp$ is an \ac{iid} zero-mean and unit-variance complex Gaussian vector.
\item $\rho$ is a non-negative real constant.
\item $u(\cdot)$ is a general penalty function with decoupling property, i.e.,
\begin{align}
u(\bv)=\sum_{n=1}^N u(v_n).
\end{align}
\item $\bz_{K \times 1}$ is \ac{iid} zero-mean complex Gaussian with variance $\sigma^2$, i.e., $\bz\sim\cnor{\sigma^2 \mI_K}$.
\item $\bs$, $\bz$ and $\mH$ are independent.
\end{enumerate}

\begin{remark}
Considering the \ac{glse} precoding scheme, one observes that the mapping from the data symbols to the transmit vector $\bx$ depends on the instantaneous value of the data symbols. In other words, for several examples of \ac{glse} precoding, the precoder does not reduce to a fixed function which only depends on the \ac{csi}. In this sense, \ac{glse} precoding is considered as a so-called \textit{symbol-level} precoding scheme \cite{spano2017symbol,alodeh2018symbol}.
\end{remark}
%
\subsection{Special Cases}
For several choices of the penalty function and support, the corresponding \ac{glse} precoders reduce to some well-known precoders. For example, when $u(\bv)=\lambda \norm{\bv}^2$, we have
\begin{enumerate}[label=(\theenumi)]
\item By setting $\setX=\setC$, the precoder reduces to the~\ac{rzf}~precoder \cite{peel2005vector} which determines $\bx$ as
\begin{align}
\bx = \sqrt{\rho} \hspace*{.7mm} \mH^\her (\mH \mH^\her + \lambda \mI_k)^{-1} \bs.
\end{align}
\item By assuming $\setX = \set{z\in\setC : \abs{z}=\sqrt{P}}$ for some non-negative real constant $P$, the precoder reduces to the constant envelope precoder considered in \cite{mohammed2013per}. The precoder in this case constructs the transmit vectors  with minimum overall distortion, such that the envelope of the signal on each antenna remains constant.
\item By considering a general $\setX\subset\setC$, the precoder reduces to the power-limited nonlinear \ac{lse} precoder introduced in \cite{sedaghat2017lse} and investigated in \cite{sedaghat2017least}. 
\end{enumerate}

\begin{remark}
Although \ac{glse} precoding considers a general support, restricting the support to the complex plane, i.e., $\setX=\setC$ does not change the set of signal constraints being addressed by \ac{glse} precoding. In fact, as the regularization term is general, one can invoke classical methods, e.g., the barrier method or penalty method \cite{boyd2004convex}, to pose restrictions on the constellation points via penalizing. The general precoding support in the formulation is hence for sake of compactness.
\end{remark}
\subsection{Performance Measures}
We intend to investigate the asymptotic performance of the \ac{glse} precoders. To this end, several metrics can be employed as the measure of performance, e.g., the \ac{mse} or the achievable ergodic rate. We consider the achievable ergodic rate per user which is defined as
\begin{align}
\mar_\av &\coloneqq \frac{1}{K}  \sum_{k=1}^K \E_\mH\set{\mar_k} \label{eq:mar_av} 
\end{align}
Here, $\mar_k$ denotes the achievable rate for user $k\in [K]$ when a \ac{glse} precoder is employed for construction of the transmit signal and $\E_\mH$ indicates expectation over the channel matrix $\mH$. Due to the nonlinearity of \ac{glse} precoders, the direct derivation of $\mar_\av$ is not trivial. We thus determine a lower bound on $\mar_\av$. 
For this aim, define the random variable $w_k(\mH)$ as
\begin{align}
w_k (\mH) \coloneqq [\mH\bx]_k-\sqrt{\rho} \hspace*{.3mm} s_k
\end{align}
for some real $\rho > 0$ and $k\in [K]$, where $[\mH\bx]_k$ denotes the $k$th entry of the $K\times 1$ vector $\mH\bx$. Consequently, the receive symbol at user $k$ is written as $y_k= \sqrt{\rho} \hspace*{.3mm} s_k + w_k (\mH) + z_k$. 
The achievable ergodic rate for user $k$ reads
\begin{align}
\E_\mH\set{\mar_k}=\rmI\prant{y_k;s_k|\mH} = \rmh \prant{s_k|\mH} - \rmh \prant{s_k | y_k,\mH}.
\end{align}
Using the equality $\rmh \prant{ s_k|y_k ,\mH} = \rmh \prant{ s_k - \rho^{-1/2} y_k|y_k ,\mH }$, one can write
\begin{subequations}
\begin{align}
\E_\mH\set{\mar_k}&= \rmh \prant{s_k|\mH} - \rmh \hspace*{.5mm} ( \frac{w_k(\mH)+z_k}{\sqrt{\rho}} | y_k,\mH) \\
&\stackrel{\dagger}{\geq} \rmh \prant{s_k} - \rmh \hspace*{.5mm} ( \frac{w_k(\mH)+z_k}{\sqrt{\rho}} |\mH )
\end{align}
\end{subequations}
where $\dagger$ is concluded due to the facts that $s_k$ is independent of $\mH$ and $\rmh \prant{ x|y} \leq \rmh \prant{x}$. Noting that $w_k(\mH)$ and $z_k$ are independent, we can further write that
\begin{align}
\rmh \hspace*{.5mm} ( \frac{w_k(\mH)+z_k}{\sqrt{\rho}}|\mH) \leq \rmh \hspace*{.5mm} ( \frac{w_k(\mH)+z_k}{\sqrt{\rho}}) \stackrel{\star}{\leq} \log \prant{ \pi e \hspace*{.5mm} \frac{ \sigma^2 + \psi_k }{\rho} } \label{eq:lower1}
\end{align}
where $\psi_k$ denotes the variance of $w_k(\mH)$, and the equality in $\star$ holds when $w_k(\mH)$ is Gaussian. Since $s_k\sim\mathcal{CN}(0,1)$, one can bound the ergodic achievable rate for user $k$ from below as
\begin{align}
\E\set{\mar_k} \geq \log\prant{\frac{\rho}{\sigma^2+\psi_k}}. \label{eq:mar_k}
\end{align}
By substituting in \eqref{eq:mar_av}, one derives the following bound on the achievable ergodic rate per user
\begin{subequations}
\begin{align}
\mar_\av &\geq \frac{1}{K} \sum_{k=1}^K \log \prant{ \frac{\rho}{\sigma^2+\psi_k}} \\ 
&\stackrel{\star}{\geq}  \log\prant{ \dfrac{\rho}{\sigma^2+K^{-1} \sum_{k=1}^K \psi_k}} \label{eq:lower}
\end{align}
\end{subequations}
where $\star$ is derived using Jensen's inequality and is tight when $\psi_k$ is constant in $k$, i.e., $\psi_k=\psi_\ell$ for all $k,\ell\in [K]$.  \eqref{eq:lower} gives a lower bound on the achievable ergodic rate in terms of the total interference at the user terminals. The term describing the interference in \eqref{eq:lower} is in fact the \ac{mse} between the noise-free version of the receive signals and the data symbols desired to be received at user terminals. We call this term the ``asymptotic input-output distortion'' and define it as following:

\begin{definition}[\bfseries Asymptotic Distortion]
\label{def:asy_dist}
Consider the data vector $\bs$ being precoded by the \ac{glse} precoder $\glse\prant{\cdot|\cdot}$ and transmitted over the \ac{mimo} channel $\mH$. For a given power control factor $\rho$, the asymptotic input-output distortion is denoted by $\sfD(\rho)$ and defined as
\begin{align}
\sfD \prant{\rho} \coloneqq\lim_{K\uparrow\infty} \frac{1}{K} \E \set{\norm{\mH \bx- \sqrt{\rho} \hspace*{.6mm}\bs}^2},
\end{align}
where $\bx=\glse\prant{\bs|\rho,\mH}$.
\end{definition}
Noting that
\begin{align}
\sfD \prant{\rho} = \frac{1}{K} \sum_{k=1}^K \psi_k, \label{eq:dist-psi}
\end{align}
we can conclude the following lemma from \eqref{eq:lower}.

\begin{lemma}
\label{lemma1}
In the large-system limit, the achievable ergodic rate per user $\mar_\av$ is bounded in terms of the asymptotic distortion $\sfD \prant{\rho}$ from below as $\mar_\av \geq \mar_\av^{\rm L}$ where
\begin{align}
\mar_\av^{\rm L} &\coloneqq \log\prant{ \dfrac{\rho}{\sigma^2+ \sfD \prant{\rho}}}. \label{eq:lower-inal}
\end{align}
\end{lemma}

In order to investigate the statistical properties of the transmit vector, we further define the ``asymptotic marginal'' of the transmit vector $\bx$ as follows.
\begin{definition}[\bfseries Asymptotic Marginal]
\label{def:margin}
Consider the real-valued function $f(\cdot)$ being defined over $\setX$, i.e., $f(\cdot): \setX \mapsto \setR$. Define the marginal of $f(\bx)$ over the index subset $\setW_N \subseteq [N]$ as
\begin{align}
\sfM_{f}^{\setW}(\bx;N)\coloneqq \frac{1}{\abs{\setW_N}} \sum_{w\in\setW_N } \E f(x_w).
\end{align}
The asymptotic marginal of $f(\bx)$ over the limit of $\setW_N$ is denoted by $\sfM^{\setW}_f(\bx)$ and is defined to be the limit of $\sfM^{\setW}_f(\bv;N)$ as $N\uparrow \infty$, i.e.,
\begin{align}
\sfM^{\setW}_f(\bx) \coloneqq \lim\limits_{N\uparrow\infty}\sfM^{\setW}_f(\bx;N).
\end{align}
\end{definition}
The asymptotic marginal describes the statistics of the transmit vector. To illustrate the functionality of this measure, let $f(x)=x^m$, and assume that $\setW_N=[n:n+U]$ for some integer $U$ and $n\in[1:N-U]$. In this case the marginal of $f(\bx)$ over $\setW_N$ reads
\begin{align}
\sfM_{f}^{\setW}(\bx;N)\coloneqq \frac{1}{1+U} \sum_{w=n}^{n+U} \E x_w^m.
\end{align}
which determines the averaged $m$th moment of the transmit entries $\set{x_n, \ldots,x_{n+U}}$. By setting $U= 0$, $\sfM_{f}^{\setW}(\bx;N)$ reduces to the $m$th moment of the entry $x_n$, and therefore, the $\sfM^{\setW}_f(\bx)$ gives the asymptotic $m$th moment of the precoder's $n$th output symbol. The asymptotic marginal enables us to determine the asymptotic marginal distribution of the transmit symbols and justify the marginal decoupling property of \ac{glse} precoders which is presented in Section \ref{sec:dec}.
%
\subsection{Stieltjes and $\rmR$-Transforms}
Before stating the main results, we need to define the Stieltjes and $\rmR$-transforms of a given probability distribution.
\begin{definition}[Stieltjes and $\rmR$-transform]
\label{def:r-trans}
Consider the random variable $t$ with distribution $\rmp_t$. The Stieltjes transform is given by 
\begin{align}
\rmG_t(s) = \E (t-s)^{-1}. \label{eq:stieltjes}
\end{align}
where $s\in\setC$ and $\img{s} > 0$.
Denoting the inverse \ac{wrt} composition by $\rmG_t^{-1} (\cdot)$, the {$\rmR$-transform} of $\rmp_t$ reads
\begin{align}
\rmR_t (\omega) \hspace*{-.7mm}=\hspace*{-.7mm} \rmG_t^{-1} (-\omega) - \frac{1}{\omega},
\end{align}
such that $\lim\limits_{\omega\downarrow 0} \rmR_t (\omega) = \E t$. Moreover, let $\mM_{N \times N}$ be decomposed as $\mM=\mU \mathbf{\Lambda} \mU^{-1}$ with $\mathbf{\Lambda}_{N \times N}$ being the diagonal matrix of eigenvalues, and $\mU_{N \times N}$ being the matrix of eigenvectors. Then, $\rmR_t(\mM)$ is an $N \times N$ matrix defined as $\mathrm{R}_t(\mM)=\mU \ \mathrm{diag}[\mathrm{R}_t(\lambda_1), \ldots, \mathrm{R}_t(\lambda_N)] \ \mU^{-1}$.
\end{definition}

In the following section, we represent the asymptotic distortion and marginal of $\bx$ in terms of the statistics of $\bs$ and $\mH$. Our derivations follow the replica method from statistical mechanics. The details of the large-system analysis are given later in Section~\ref{sec:large}.

\section{Asymptotics of \ac{glse} Precoders}
\label{sec:result}

To derive asymptotics of \ac{glse} precoders, we invoke the replica method from statistical mechanics which has been accepted as an analytic tool in multiuser communications and information theory. The validity of the solution determined via the replica method relies on the conjecture of ``replica continuity'' taken prior to derivations. Although this conjecture misses a generic rigorous justification, analytical and numerical investigations in the literature have approved its validity for several problems \cite{reeves2016replica,barbier2017stochastic}. The replica continuity conjecture is illustrated explicitly in Section~\ref{sec:large}.

Assuming replica continuity, the asymptotic distortion is calculated in terms of the solution of a matrix-valued fixed-point equation. This generic solution is referred to as the ``general replica solution'' which is of a complicated form. The solution however becomes of a simple form when the problem exhibits the so-called \ac{rs}. In fact, for some choices of the penalty function and precoding support, the fixed-point equation in the general replica solution shows some forms of symmetry. As the result, the solution to the fixed-point equation is found by a simple search over these symmetric matrices. Such a solution is referred to as the \ac{rs} solution in the literature. For a large scope of quadratic optimization problems, solved by \ac{glse} precoders, the \ac{rs} solution is known to be valid, meaning that the problems exhibit \ac{rs} \cite{reeves2016replica,barbier2017stochastic}. There are however some particular examples in which this symmetry does not hold \cite{zaidel2012vector,bereyhi2016statistical,sedaghat2017least}. For these cases, the classical approach is to recursively deviate from this symmetry via the \ac{rsb} scheme \cite{parisi1980sequence}. The solution in this case is referred to as the \ac{rsb} solution. 

In this section, we state the \ac{rs} solution, as well as the \ac{rsb} solution with one step of recursion. For sake of compactness, we skip the derivations here and give them in Section~\ref{sec:large} along with the~large-system analysis. Extension to the \ac{rsb} solutions with more steps of recursion is also discussed in Appendix~\ref{app:b-rsb}.

\subsection{The \ac{rs} Solution}
\label{sec:rs}
Proposition \ref{thm:1} represents the asymptotic distortion, as well as the asymptotic marginal of the transmit signal, under the \ac{rs} assumption. The \ac{rs} assumption is explicitly stated in Section \ref{sec:large}. 
\begin{proposition}
\label{thm:1}
Assume that the \ac{rs} assumption holds. Let $s^\rs\sim\mathcal{CN}(0,\rho^\rs)$ in which 
\begin{align}
\rho^\rs=\left[\rmR_\mJ(-\chi)\right]^{-2}\frac{\partial}{\partial \chi} \left[ ( \rho \chi- \sfp ) \rmR_\mJ(-\chi) \right]
\end{align}
for some $\chi$ and $\sfp$ where $\rmR_\mJ(-\chi)$ denotes the $\rmR$-transform of $\rmF_\mJ$. Define 
\begin{align}
\xx=\argmin_{v\in \setX} \abs{v- s^\rs}^2+ \left[\rmR_\mJ(-\chi)\right]^{-1} u(v). \label{eq:snigle}
\end{align}
Then, the asymptotic marginal of $f(\bx)$ is given by
\begin{align}
\sfM^{\setW}_f(\bx)= \E f(\xx), \label{eq:rs-marginal}
\end{align}
and the asymptotic distortion reads
\begin{align}
\sfD_{\rm RS} \prant{\rho}=\rho+\alpha^{-1} \frac{\partial}{\partial \chi} \left[ ( \sfp -\rho \chi ) \chi \rmR_\mJ(-\chi) \right] \label{eq:rs-dist}
\end{align}
for $\sfp$ and $\chi$ which are determined from the following set of fixed point equations
\begin{subequations}
\begin{align}
\sfp &= \E \abs{\xx}^2 \label{eq:fix-rs1} \\
\chi \rmR_\mJ (-\chi) &= \frac{1}{\rho^\rs} \E \re{\xx^* s^\rs} . \label{eq:fix-rs2}
\end{align}
\end{subequations}
In the case that the solution to \eqref{eq:fix-rs1} and \eqref{eq:fix-rs1} is not unique, the fixed-point is chosen such that $\sfD_{\rm RS}$ is minimized. 
\end{proposition}
\begin{prf}
The proof is given in Section \ref{sec:large}.
\end{prf}


Via numerical investigations, we later show that the \ac{rs} solution is exact in various cases. It is however demonstrated that in some particular settings $\sfD_{\rm RS} \prant{\rho}$ gives a loose lower bound on $\sfD \prant{\rho}$ meaning that the problem in these cases does not show \ac{rs}. For those scenarios, we express the solution under one-step~\ac{rsb}.

\subsection{The One-step \ac{rsb} Solution}
\label{sec:rsb}
Several studies in the literature have shown examples in which the \ac{rs} solution fails to track the asymptotic behavior of the system; see for example \cite{zaidel2012vector,bereyhi2016statistical}. Regarding \ac{glse} precoding, one can guarantee the existence of such particular examples from the investigations in \cite{sedaghat2017lse}. In fact, for nonlinear \ac{lse} precoding in \cite{sedaghat2017lse}, which is a special form of \ac{glse} precoding, it has been demonstrated that the \ac{rs} solution violates known lower bounds in some cases. From the literature, it is known that the accurate asymptotic performance in these cases is addressed by modifying the \ac{rs} assumption via the recursive \ac{rsb} scheme \cite{parisi1980sequence}. The details on \ac{rsb} are given in Section~\ref{sec:large}. In general, the \ac{rsb} scheme can be recurred for multiple number of steps. Proposition~\ref{thm:2} represents the result for one-step of \ac{rsb}. Extension to more steps is later discussed in Appendix \ref{app:b-rsb}.

\begin{proposition}
\label{thm:2}
Consider the \ac{rsb} scheme with one step of recursion. For some given $\chi$, $\sfp$, $\mu$ and $\sfc$, define $\rho^\rs$ and $\rho^\rsb_1$ as
\begin{subequations}
\begin{align}
\rho^\rs&=[\rmR_\mJ(-\chi)]^{-2}\frac{\partial}{\partial \tilde{\chi}} \left[ ( \rho \tilde{\chi}- \sfp ) \rmR_\mJ(-\tilde{\chi}) \right]\\
\rho^\rsb_1&=[\rmR_\mJ(-\chi)]^{-2} \mu^{-1} \left[ \rmR_\mJ(-\chi)-\rmR_\mJ(-\tilde{\chi}) \right] \label{eq:rho1}
\end{align}
\end{subequations}
where $\tilde{\chi}\coloneqq\chi+\mu\sfc$. Let 
\begin{align}
\xx=\argmin_{v\in\setX} \abs{v-  s^\rs - s^\rsb_1}^2+ [\rmR_\mJ(-\chi)]^{-1} u(v). \label{eq:single}
\end{align}
where $s^\rs \sim \mathcal{CN}(0;\rho^\rs)$, and $s^\rsb_1$ is distributed conditional to $s^\rs$ with
\begin{align}
\rmp_1^\rsb(u|t)  = \dfrac{\exp\set{-{\mu \rmR_\mJ(-\chi)} \left[ \abs{\xx-  u - t}^2-\abs{ u + t}^2 \right] -\mu u(\xx)} \phi(u;\rho_1^\rsb)}{\int_{\setC} \exp\set{-{\mu \rmR_\mJ(-\chi)} \left[ \abs{\xx-  w - t}^2-\abs{ w + t}^2 \right] -\mu u(\xx)} \phi(w;\rho_1^\rsb) \dif w}. \label{final_prob}
\end{align}
%
Then, the marginal distribution reads
\begin{align}
\sfM^{\setW}_f(\bx)= \E f(\xx),
\end{align}
and the distortion asymptotically tends to
\begin{align}
\sfD_{\rm RSB}(\rho)=\rho  + \alpha^{-1} \left\lbrace \frac{\partial}{\partial \tilde{\chi}}\left[ ( \sfp-\rho \tilde{\chi} ) \tilde{\chi} \rmR_\mJ(-\tilde{\chi}) \right]  + \frac{\xi\sfp  -\tilde{\chi}\rho_1^\rsb}{\xi^2} \right\rbrace, \label{final_dist}
\end{align}
for ${\chi}$, $\sfc$ and $\sfp$ which are determined via the fixed-point equations
\begin{subequations}
\begin{align}
\sfc+\sfp&=\E \abs{\xx}^2\\
\sfp+\tilde{\chi} &= \frac{\xi}{\rho_1^\rsb} \hspace*{.5mm} \E \re{\xx^* s_1^\rsb}  \label{eq:fix2} \\
\tilde{\chi} &= \frac{\xi}{\rho^\rs} \hspace*{.5mm} \E \re{\xx^* s^\rs}.
\end{align}
\end{subequations}
and $\mu$ which satisfies 
\begin{align}
\frac{\mu^2 \sfp}{\xi^2} \rho_1^\rsb + \frac{\mu \sfc}{\xi} -\int_{\chi}^{\tilde{\chi}} \rmR_{\mJ}(-\omega) \dif \omega =\mathrm{I} \left( s_1^\rsb;s^\rs \right) + \mathrm{D}_{\mathsf{KL}} ( \rmp_{s_1^\rsb} \Vert \phi(\cdot;\rho_1^\rsb) ). \label{eq:fix3}
\end{align}
Here, $\mathrm{I} \left( s_1^\rsb;s^\rs \right)$ represents the mutual information between $s^\rs$ and $s^\rsb_1$. Moreover, $\rmp_{s_1^\rsb}$ indicates the marginal distribution of $s_1^\rsb$ determined by
\begin{align}
\rmp_{s_1^\rsb}(u)= \int \rmp_1^\rsb (u|t) \phi(t;\rho^\rs) \dif t.
\end{align}
$\mathrm{D}_{\mathsf{KL}}( \rmp_{s_1^\rsb} \Vert \phi(\cdot;\rho_1^\rsb) )$ denotes the Kullback–Leibler divergence between $\rmp_{s_1^\rsb}$ and a zero-mean complex Gaussian distribution with variance $\rho_1^\rsb$ defined as
\begin{align}
\mathrm{D}_{\mathsf{KL}}( \rmp_{s_1^\rsb} \Vert \phi(\cdot;\rho_1^\rsb) ) \coloneqq \int \rmp_{s_1^\rsb}(u) \log \frac{\rmp_{s_1^\rsb}(u)}{\phi(u;\rho_1^\rsb)} \dif u.
\end{align}
In the case that the fixed-point equations have multiple solutions, $\chi$, $\sfp$, $\sfc$ and $\mu$ are chosen such that $\sfD_{\rm RSB}(\rho)$ is minimized.
\end{proposition}

\begin{prf}
The detailed derivations are given in Section \ref{sec:large}.
\end{prf}

\subsection{Discussions on the \ac{rs} and \ac{rsb} Solutions}
\label{sec:discussion}
There are several approaches to investigate whether the solution given under \ac{rs} is tight or further investigations under \ac{rsb} are required. Discussing of these approaches is out of the scope of this study and can be followed in \cite{mezard2009information} and the references therein. The available results in the literature, however, can be employed to check the tightness of the replica solutions. There is a well-known belief that the asymptotic solution to convex optimization problems is precisely recovered via the \ac{rs} solution. The belief has been discussed for some particular examples in the literature \cite{moustakas2007outage}. Based on these discussions, it is conjectured that for \ac{glse} precoders with convex penalty and support, the \ac{rs} solution is exact. 

A basic approach to test the existence of \ac{rs} is the so-called zero-temperature entropy test. In this test, the entropy of the corresponding thermodynamic system\footnote{The concepts of entropy and temperature are defined in terms of the corresponding thermodynamic system. See \cite{bereyhi2016statistical} for more details.} is determined, and its convergence is checked when the temperature tends to zero. When the replica solution is perfectly tracking the performance, the zero-temperature entropy should converge to zero. However, for cases in which the \ac{rs} or one-step \ac{rsb} solution does not give the exact solution, the respective zero-temperature entropy does not converge to zero. For these cases, it is shown that the exact asymptotic distortion is bounded from below by both the \ac{rs} and one-step \ac{rsb} solutions, and the bound given by the on-step \ac{rsb} is tighter than the \ac{rs} one. By further recursion steps of \ac{rsb}, the lower bound given by the \ac{rsb} solution becomes tighter. In this case, the \ac{rsb} solution with an infinite number of recursions meets the exact asymptotic distortion\footnote{The solution in this case is called the full \ac{rsb} solution. There are also some cases in which the solution becomes exact with finite steps of \ac{rsb} recursions \cite{talagrand2010mean}.}\cite{guerra2003broken,talagrand2006parisi}. Throughout the numerical investigations of the results in Section~\ref{sec:application}, we show that for most \ac{glse} precoders, the \ac{rs} solution tracks the exact performance consistently and there are few cases for which we need further \ac{rsb} investigations. For the latter cases, the one-step \ac{rsb} solution is shown to give a tight lower bound for a large regime of loads, i.e., $\alpha=K/N$.

\subsection{Marginal Decoupling Property}
\label{sec:dec}
Propositions~\ref{thm:1}~and~\ref{thm:2} indicate that the asymptotic marginal of $\bx$ has the following properties:
\begin{enumerate}[label=(\alph*)]
\item For any function $f(\cdot)$, it does not depend on the index set $\setW$.
\item It is equal to the expected marginal of an equivalent scalar \ac{glse} precoder which reads
\begin{align}
\glse_\dec (s^\dec|\xi) \coloneqq \argmin_{v\in\setX} \abs{v-  s^\dec}^2+ \xi \hspace*{.5mm} u(v) \label{eq:dec}
\end{align}
for $\xi = [\rmR_\mJ(-\chi)]^{-1}$. The random variable $s^\dec$ is moreover given by $s^\dec =  s^\rs$ under \ac{rs} and $s^\dec  =  s^\rs+s_1^\rsb$ under one-step \ac{rsb}.
\end{enumerate}
These findings state that the statistics of $\xx=\glse_\dec (s^\dec|\xi)$ describe the asymptotic statistical properties of the output entries. More precisely, one can conclude from these statements that all the output entries of $\bx$ are asymptotically identically distributed with a marginal distribution equal to the distribution of $\xx$. We refer to this phenomenon as the ``marginal decoupling property''.

%
%
%

\begin{proposition}[Marginal Decoupling Property]
\label{thm:3}
Assume that the asymptotic distortion of \ac{glse} precoding is given by Proposition~\ref{thm:1}~or~\ref{thm:2}. Then, the transmit symbol $x_n$ for $n\in[N]$ converges in distribution to the random variable 
\begin{align*}
\xx=\glse_\dec (s^\dec|\xi)
\end{align*}
as $N$ grows large, where $\glse_\dec (\cdot|\xi)$ is defined in \eqref{eq:dec}, and $s^\dec = s^\rs$ under \ac{rs} and $s^\dec = s^\rs+s_1^\rsb$ under one-step \ac{rsb}.
\end{proposition}
\begin{prf}
The proof directly follows the replica results given in Propositions \ref{thm:1} and \ref{thm:2} by employing methods from the classical moment problem \cite{akhiezer1965classical}. Without loss of generality, we consider the \ac{rs} solution. Starting from Proposition \ref{thm:1}, let $f(x)=x^m$ for $m\in\setZ$, and consider the index set $\setW_N=[n:n+\zeta N]$ for some $\zeta\in [0,1]$~and~$n\in[N]$. Substituting in Definition \ref{def:margin}, we have 
\begin{align}
\sfM_{f}^{\setW}(\bx;N)&= \frac{1}{1+\zeta N} \sum_{w=n}^{n+\zeta N} \E x_w^m.
\end{align}
$\sfM_{f}^{\setW}(\bx;N)$ determines the arithmetic average of the $m$th moment of the output entries indexed by the set $\setW_N$. By taking the limit $N\uparrow\infty$, we have
\begin{subequations}
\begin{align}
\sfM_{f}^{\setW}(\bx)&= \lim_{N\uparrow\infty} \frac{1}{1+\zeta N} \sum_{w=n}^{n+\zeta N} \int x_w^m \rmp_w^N(x_w) \dif x_w\\
&=\lim_{N\uparrow\infty} \frac{1}{1+\zeta N} \sum_{w=n}^{n+\zeta N} \int x_w^m \rmp_w(x_w) \dif x_w \label{eq:marg1}
\end{align}
\end{subequations}
where $\rmp_w^N$ indicates the distribution of the $w$th entry of $\bx\in\setC^{N}$ and $\rmp_w$ denotes its asymptotic limit. On the other hand, using Proposition \ref{thm:1}, $\sfM_{f}^{\setW}(\bx)$ is given by
\begin{align}
\sfM_{f}^{\setW}(\bx)&= \E \xx^m \label{eq:marg2}
\end{align}
which is the $m$th moment of the random variable $\xx$. The \ac{rhs} of \eqref{eq:marg2} is not a function of $\zeta$ which indicates that it holds for any choice of $\zeta$. Therefore, by setting the \ac{rhs} of \eqref{eq:marg1} and \eqref{eq:marg2} to be equal, we have
\begin{align}
\lim_{N\uparrow\infty} \frac{1}{1+\zeta N} \sum_{w=n}^{n+\zeta N} \int x_w^m \rmp_w(x_w) \dif x_w= \E \xx^m \label{eq:marg3}.
\end{align}
Now by letting $\zeta\downarrow 0$, \eqref{eq:marg2} reduces to
\begin{align}
\int x_n^m \rmp_n(x_n) \dif x_n= \E \xx^m, \label{eq:marg4}
\end{align}
for any $m\in\setZ$. \eqref{eq:marg4} along with the classical moment method concludes the proof. In fact, as the moments of $x_n$ and $\xx$ are uniformly bounded and equal for any $m\in\setZ$, one can use Carleman's Theorem from the moments method \cite{akhiezer1965classical}, and conclude that the distribution of these two random variables should be the same as well. The arguments here hold for any $n\in [N]$ which let us further conclude that for any index $n$, $\rmp_n$ tends to the distribution of $\xx$.


\end{prf}

Throughout the manuscript, we call $\xx$ the ``decoupled transmit'' and $s^\dec$ the ``decoupled input'' symbol. $\glse_\dec(\cdot|\cdot)$ is moreover referred to as the decoupled \ac{glse} precoder. One should note that although the decoupling property determines the marginal distributions of transmit entries $x_n$, it does not describe the correlation among these entries.

\section{Applications of \ac{glse} Precoders}
\label{sec:application}
The generality of the penalty function and precoding support in \ac{glse} precoders lets us consider several transmission constraints at massive \ac{mimo} base stations. Moreover, various well-known precodings, whose exact performances are not completely known in the literature, can be considered as special cases of \ac{glse} precoding. In this section, we study several forms of the \ac{glse} precoders which address various restrictions on the transmit signal, namely scenarios with a limited number of transmit \ac{rf}-chains, restricted \ac{papr} and discrete constellations. We then employ the asymptotic results presented in the previous section to investigate the large-system performance of these precoders.

As the results enclose the class of unitarily invariant channel matrices, the performance of \ac{glse} precoding can be investigated for several fading models. Throughout the investigations, we consider a standard \ac{iid} flat Rayleigh fading channel whose channel matrix in the downlink~reads
\begin{align}
\mH= \mA^{1/2} \mG \label{eq:channel}
\end{align}
where $\mG_{K\times N}$ contains \ac{iid} zero-mean Gaussian entries with variance $1/N$, and $\mA_{K\times K}=\mathrm{diag}\left[ a_1, \ldots, a_K \right]$ is a diagonal matrix with $a_k$ being non-negative real variables for $k\in[K]$. Here, the matrix $\mG$ models the multipath effect and $\mA$ describes the path-loss and shadowing. The diagonal entries of $\mA$ are considered to be random variables whose mean values depend on the positions of the users in the network. We denote the limiting empirical cumulative distribution of $\set{a_1, \ldots, a_K}$ by $\rmF^{\sf snr}$.

To employ the asymptotic results, the $\rmR$-transform of $\mJ=\mH^\her \mH$ is required to be derived. By considering the channel model in \eqref{eq:channel}, one notes that $\mA^\her=\mA$ and writes
\begin{align}
\mJ=\mG^\her \mA \mG. \label{eq:Matrix}
\end{align}
For the case of $\mA=\mI_K$, the asymptotic distribution $\rmF_\mJ$ follows the Mar{\u c}enko-Pastur law which implies that the $\rmR$-transform in this case is given by \cite{muller2013applications,tulino2004random}
\begin{align}
\rmR_\mJ(\omega)=\frac{\alpha}{1-\omega}. \label{eq:mat-iid}
\end{align}
When $\mA\neq \mI_K$, $\rmF_\mJ$ is derived from the results in \cite{silverstein1995empirical}\footnote{This result is given in \cite[Section 3.2.2]{muller2013applications}.} which leads to
\begin{align}
\rmR_\mJ(\omega)= \alpha \int \frac{a}{1-a\omega} \dif \rmF^{\sf snr}(a). \label{eq:shadow}
\end{align}
For sake of brevity, we assume $\mA=\mI_K$ for our analytical investigations. The results~straightforwardly extend to cases with $\mA\neq \mI_K$, by substituting \eqref{eq:shadow} into the derivations. 

In the sequel, we study the asymptotic performance of \ac{glse} precoding for multiple constraints on the transmit signal. Initially, we derive the \ac{rs} solution and discuss its validity via numerical investigations. The analysis is later extended to the one-step \ac{rsb} solution, if the \ac{rs} solution fails to give a tight bound on the asymptotic performance.

\subsection{Optimal Antenna Selection}
\label{sec:ex1}

Assume that the precoded symbols can be taken from the whole complex plane, i.e., $\setX=\setC$. This assumption means that the transmit signal is not restricted in terms of \ac{papr}\footnote{The extension into a case with \ac{papr} restriction is discussed later}. In this section, we follow the discussions in Section~\ref{sec:AntSelect} and simultaneously restrict the number of active transmit antennas and the transmit power via \ac{glse} precoding. For this aim, we set 
\begin{align}
u(\bv)=\lambda \norm{\bv}^2 + \lambda_0 \norm{\bv}_0. \label{eq:penalty0}
\end{align}
The first term in \eqref{eq:penalty0} controls the transmit power, i.e., $\norm{\bx}^2 / N$, and the second term restricts the fraction of active transmit antennas, i.e., $\norm{\bx}_0 / N$. In this case, a pair of constraints on the transmit power and the number of active antennas are satisfied by tuning $\lambda$ and $\lambda_0$, correspondingly. We will discuss this tuning problem within the derivations.

Considering the asymptotic distortion as the metric which measures the performance of the system, the \ac{glse} precoder with penalty in \eqref{eq:penalty0} is the optimal algorithm for joint power control and \ac{tas}, since it solves the corresponding constrained optimization problem. The algorithm is however computationally complex, since the number of searches, needed for finding $\bx$, grows exponentially large \ac{wrt} $N$. An effective algorithm is obtained by relaxing the $\ell_0$-norm term in \eqref{eq:penalty0} by $\ell_1$-norm. We discuss this alternative algorithm later on. 
\begin{remark}
One should note that in \ac{glse} precoders the signal constraints are satisfied instantaneously. Considering \ac{tas}, this means that active transmit antennas are selected by \ac{glse} precoders in each transmission interval. In practical settings, one might be interested in some approaches in which active antennas are selected on a lower rate. This issue can be further addressed by a standard extension of the current \ac{glse} precoding scheme such that the precoder constructs a frame of transmit signals from a block of data symbols jointly. Further discussions on this extension is presented in Appendix~\ref{app:block_GLSE}. 
\end{remark}

\subsubsection*{Asymptotic Distortion}
Using Proposition~\ref{thm:1}, the decoupled \ac{glse} precoder in this case reads 
\begin{align}
\glse_\dec(s^\rs|\xi)=
\begin{cases}
    \dfrac{s^\rs}{1+\xi\lambda} \qquad &\abs{s^\rs}\geq \tau_0 \\
    0             &\abs{s^\rs} < \tau_0 \label{eq:sing0}
\end{cases}
\end{align}
for $\xi\coloneqq\alpha^{-1}(1 + \chi)$ where $s^\rs \sim \mathcal{CN}(0, \rho^\rs)$ and the threshold $\tau_0$ is given by
\begin{align}
\tau_0 \coloneqq\sqrt{\xi\lambda_0 (1+\xi \lambda)}.
\end{align}
Moreover, by invoking \eqref{eq:mat-iid}, $\rho^\rs$ for this setup is given by 
\begin{align}
\rho^\rs=\alpha^{-1}(\rho + \sfp). \label{eq:rho_RS_AA}
\end{align}
By determining the asymptotic marginal of the transmit signal from Proposition~\ref{thm:1} for $f(x)=\abs{x}^2$, it is observed that $\sfp$ represents the average transmit power, i.e. 
\begin{align}
\sfp=\lim_{N\uparrow\infty}\E\frac{1}{N} \norm{\bx}^2 ,
\end{align}
which reads
\begin{align}
\sfp= \dfrac{\rho^\rs + \tau_0^2}{\left(1+\xi\lambda \right)^2} \exp\set{-\frac{\tau_0^2}{\rho^\rs}}, \label{eq:sfp}
\end{align}
and $\chi\in\setR^+$ satisfies the following fixed-point equation which is coupled with \eqref{eq:sfp}
\begin{align}
{\rho^\rs} \chi&= {\xi \sfp +\xi^2 \lambda \sfp}. \label{eq:chi0}
\end{align}
The decoupled setting in \eqref{eq:sing0} indicates that the decoupled transmit signal $\xx$ is obtained by hard thresholding of the decoupled input $s_\rs$. The threshold level $\tau_0$ depends on $\lambda_0$. By setting $\lambda_0=0$, the threshold reduces to zero, and therefore, the \ac{glse} precoder under study describes the output distribution of the transmit signal constructed via the \ac{rzf} precoder; see \cite{sedaghat2017least}. 

The asymptotic distortion in this case is calculated by
\begin{align}
\sfD_{\rm RS}(\rho) =\frac{\rho+\sfp}{\left(1+\chi\right)^2}. \label{eq:sfD0}
\end{align}
for $\sfp$ and $\chi$ fulfilling \eqref{eq:sfp} and \eqref{eq:chi0}. 
\subsubsection*{Tuning $\lambda$ and $\lambda_0$}
To satisfy a desired pair of constraints on the transmit signal, $\lambda$ and $\lambda_0$ need to be tuned. To this end, we note that the asymptotic average transmit power for given $\lambda$ and $\lambda_0$ is determined in \eqref{eq:sfp}. Moreover, the asymptotic fraction of active antennas, defined as\footnote{Here, $\norm{\bx}_0$ counts the number of non-zero entries of $\bx$, and thus, represents the number of active antennas.}
\begin{align}
\eta \coloneqq \lim_{N\uparrow\infty} \frac{1}{N} \norm{\bx}_0,
\end{align}
is calculated from the asymptotic marginal of the transmit signal for the indicator function, i.e. $f(x)=\mone\set{x\neq 0}$ which leads to
\begin{align}
\sfa = \exp\set{-\frac{\tau_0^2}{\rho^\rs}}. \label{eq:sfa}
\end{align}
In systems with large dimensions, $\lambda$ and $\lambda_0$ are tuned using \eqref{eq:sfa} and \eqref{eq:sfp} as follows: Assume that the transmit power and the fraction of active transmit antennas are desired to be $P_0$ and $\eta_0$, respectively. In this case, one can find the corresponding $\lambda$ and $\lambda_0$ from the coupled equations in \eqref{eq:sfp}, \eqref{eq:chi0} and \eqref{eq:sfa} by setting $\sfp$ and $\eta$ equal to $P_0$ and $\eta_0$, respectively.

\begin{figure}[t]
\centering
\resizebox{.94\linewidth}{!}{
\pstool[width=.35\linewidth]{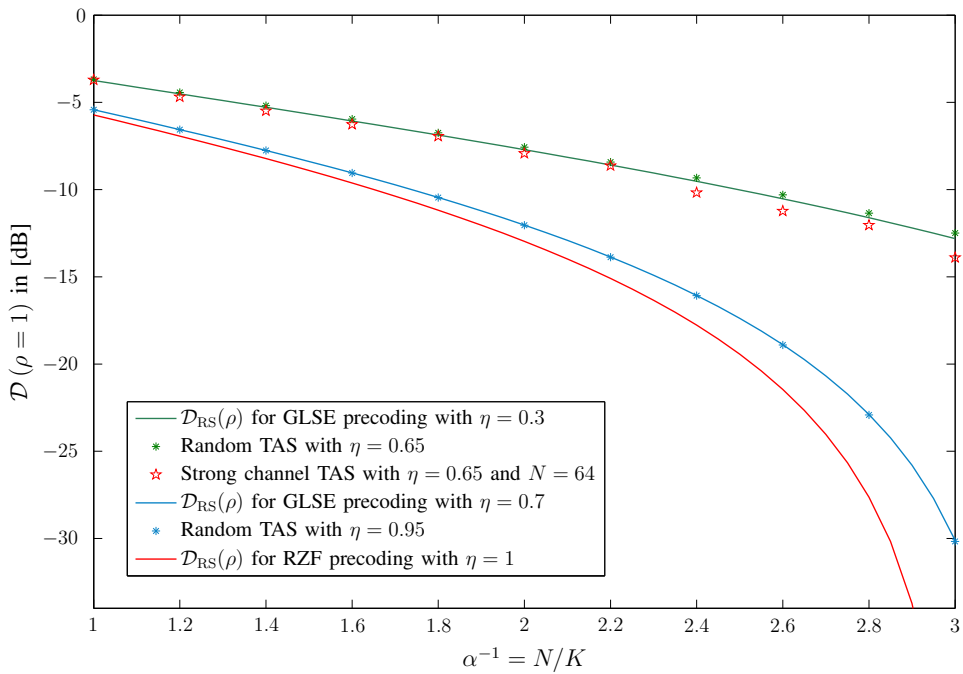}{
\psfrag{D in dB}[c][c][0.3]{$\sfD\prant{\rho=1}$ in [dB]}
\psfrag{alpha-inv}[c][b][0.3]{$\alpha^{-1}=N/K$}
\psfrag{AAABBBCCCRRREEE030DDDDSSSSBBBBA}[l][l][0.26]{$\sfD_{\rm RS}(\rho)$ for \ac{glse} precoding with $\eta=0.3$}
\psfrag{AAABBBCCCRRRNNN065DDDDSSSSBBBBA}[l][l][0.26]{Random \ac{tas} with $\eta=0.65$}
\psfrag{AAABBBN064STRNNN065DDDDSSSSBBBBA}[l][l][0.26]{Strong channel \ac{tas} with $\eta=0.65$ and $N=64$}
\psfrag{AAABBBCCCRRREEE070DDDDSSSSBBBBA}[l][l][0.26]{$\sfD_{\rm RS}(\rho)$ for \ac{glse} precoding with $\eta=0.7$}
\psfrag{AAABBBCCCRRRNNN095DDDDSSSSBBBBA}[l][l][0.26]{Random \ac{tas} with $\eta=0.95$}
\psfrag{AAABBBCCCRRREEE100DDDDSSSSBBBBA}[l][l][0.26]{$\sfD_{\rm RS}(\rho)$ for RZF precoding with $\eta=1$}


\psfrag{-5}[r][c][0.25]{$-5$}
\psfrag{-10}[r][c][0.25]{$-10$}
\psfrag{-15}[r][c][0.25]{$-15$}
\psfrag{-20}[r][c][0.25]{$-20$}
\psfrag{-25}[r][c][0.25]{$-25$}
\psfrag{-30}[r][c][0.25]{$-30$}
\psfrag{0}[r][c][0.25]{$0$}
%
\psfrag{1}[c][b][0.25]{$1$}
\psfrag{1.2}[c][b][0.25]{$1.2$}
\psfrag{1.4}[c][b][0.25]{$1.4$}
\psfrag{1.6}[c][b][0.25]{$1.6$}
\psfrag{1.8}[c][b][0.25]{$1.8$}
\psfrag{2}[c][b][0.25]{$2$}
\psfrag{2.2}[c][b][0.25]{$2.2$}
\psfrag{2.4}[c][b][0.25]{$2.4$}
\psfrag{2.6}[c][b][0.25]{$2.6$}
\psfrag{2.8}[c][b][0.25]{$2.8$}
\psfrag{3}[c][b][0.25]{$3$}

}}
\caption{Asymptotic distortion versus the inverse load $\alpha^{-1}=N/K$ for average transmit power $\sfp=0.5$ and multiple \ac{tas} constraints. The solid lines show the \ac{rs} solutions. For random and strong channel \ac{tas}, the subset of the active transmit antennas are first selected via the corresponding selection algorithm and then precoded via the \ac{rzf} precoder. The \ac{glse} precoder with $\ell_0$-norm penalty is not possible to be numerically simulated for $\eta\neq 1$, since its computational complexity grows exponentially with $N$. The result for \ac{rzf} precoding, i.e. $\eta=1$, is consistent with simulations.}
\label{fig:1}
\end{figure}

\subsubsection*{Numerical investigations}
Fig.~\ref{fig:1} shows the asymptotic distortion in terms of the inverse load, i.e. $\alpha^{-1}=N/K$, for multiple asymptotic fractions of active antennas. The control factors $\lambda$ and $\lambda_0$ are tuned such that $\sfp=0.5$, and the power control factor is set to $\rho=1$. As a benchmark, we have also plotted the distortion for a well-known \ac{tas} algorithm based on sorting the channel gains. This algorithm is illustrated in Appendix~\ref{app:tas_alg}. When the number of active antennas grows large linearly with the total number of transmit antennas, this benchmark algorithm performs close to random selection. This is observed in Fig.~\ref{fig:1} where the algorithm has been simulated for $N=64$ transmit antennas. We therefore consider the asymptotic of random selection\footnote{By random selection, in this case, we mean that the transmitter selects the subset of active antennas randomly and precodes over them using the \ac{rzf} precoding scheme.} as the reference and compare the \ac{glse} precoder against it.

As the figure depicts, for a given constraint on $\eta$, the \ac{glse} precoder significantly outperforms the \ac{rzf} precoder with random \ac{tas}.  To quantify the improvement, we have plotted the asymptotic distortion for the \ac{rzf} precoder with random \ac{tas} considering several values of $\eta$. The numerical investigations show that the performance of the \ac{glse} precoder with $\eta=0.3$ is tracked by random \ac{tas} when $65\%$ of antennas are set active. Therefore, the proposed \ac{glse} precoder with $\ell_0$-norm penalty needs around $0.35N$ less active antennas than random \ac{tas} which means $54\%$ of reduction in the number of active antennas. This improvement reduces to $0.25N$ at $\eta=0.7$. As indicated earlier, the significant improvement in the performance in this case is achieved at the expense of high computational complexity. We address this issue in the following section by replacing the $\ell_0$-norm regularization term with an $\ell_1$-norm penalty.

\subsection{Antenna Selection via Convex Penalty}
\label{sec:ex2}
From the literature of compressive sensing \cite{donoho2006compressed,candes2006robust,foucart2013mathematical}, it is known that an effective computationally feasible approach for sparse recovery is LASSO regression \cite{tibshirani1996regression}. In this approach, the sparsity of the signal is approximated via the $\ell_1$-norm leading to a convex recovery algorithm which is solved within polynomial time. Invoking this relaxation, one can restrict the number of active transmit antennas, as well as the transmit power, via \ac{glse} precoding by setting
\begin{align}
u(\bv)=\lambda \norm{\bv}^2 + \lambda_1 \norm{\bv}_1. \label{eq:penalty_ell_1}
\end{align}
For convex choices of $\setX$, this \ac{glse} precoder solves a convex optimization problem,~and~therefore, is posed by linear programming. 

To investigate the performance of this \ac{glse} precoder, we start by considering the case with no \ac{papr} restriction, i.e. $\setX=\setC$. In this case, the decoupled \ac{glse} precoder is given by
\begin{align}
\glse_\dec(s^\rs|\xi)=
\begin{cases}
   \dfrac{s^\rs}{1+\xi\lambda} \dfrac{\abs{s^\rs}-\tau_1}{\abs{s^\rs}} &\abs{s^\rs}\geq \tau_1 \\
    0             &\abs{s^\rs} < \tau_1 \label{eq:single_norm1}
\end{cases}
\end{align}
where $\tau_1\coloneqq {\xi \lambda_1}/{2}$ for $\xi\coloneqq\alpha^{-1}(1 + \chi)$ and the decoupled input reads $s^\rs\sim\mathcal{CN}\left(0,\rho^\rs\right)$ with $\rho^\rs$ given in \eqref{eq:rho_RS_AA}. The average transmit power $\sfp$ in this case is determined by
\begin{align}
\sfp=\dfrac{\rho^\rs}{\left(1+\xi\lambda \right)^2} \exp\set{-\frac{\tau_1^2}{\rho^\rs}}-2 \dfrac{\tau_1\sqrt{\pi \rho^\rs}}{\left(1+\xi\lambda \right)^2} \rmQ ( \sqrt{\frac{2}{\rho^\rs}} \tau_1 )
\end{align}
with $\rmQ(\cdot)$ denoting the standard $\rmQ$-function. The asymptotic fraction of active antennas reads
\begin{align}
\eta = \exp\set{-\frac{\tau_1^2}{\rho^\rs}}. \label{eq:sfa1}
\end{align}
Moreover, the scalar $\chi$ is found by solving the fixed-point equation 
\begin{align}
\frac{\alpha \chi}{1+\chi} \left(1+\xi\lambda\right)= \exp\set{-\frac{\tau_1^2}{\rho^\rs}}- \dfrac{\tau_1\sqrt{\pi \rho^\rs}}{\rho^\rs } \rmQ ( \sqrt{\frac{2}{\rho^\rs}} \tau_1 ).
\end{align}
Considering the decoupled \ac{glse} precoder in \eqref{eq:single_norm1}, one observes that under $\ell_1$-norm minimization the decoupled precoder is a soft thresholding operator which reduces to a linear precoder as $\lambda_1$ tends to zero. For given constraints on the transmit power and the fraction of active antennas, the tuning strategy follows the same approach as in Section \ref{sec:ex1}.

\subsubsection*{Numerical investigations}
\begin{figure}[t]
\centering
\resizebox{.94\linewidth}{!}{
\pstool[width=.35\linewidth]{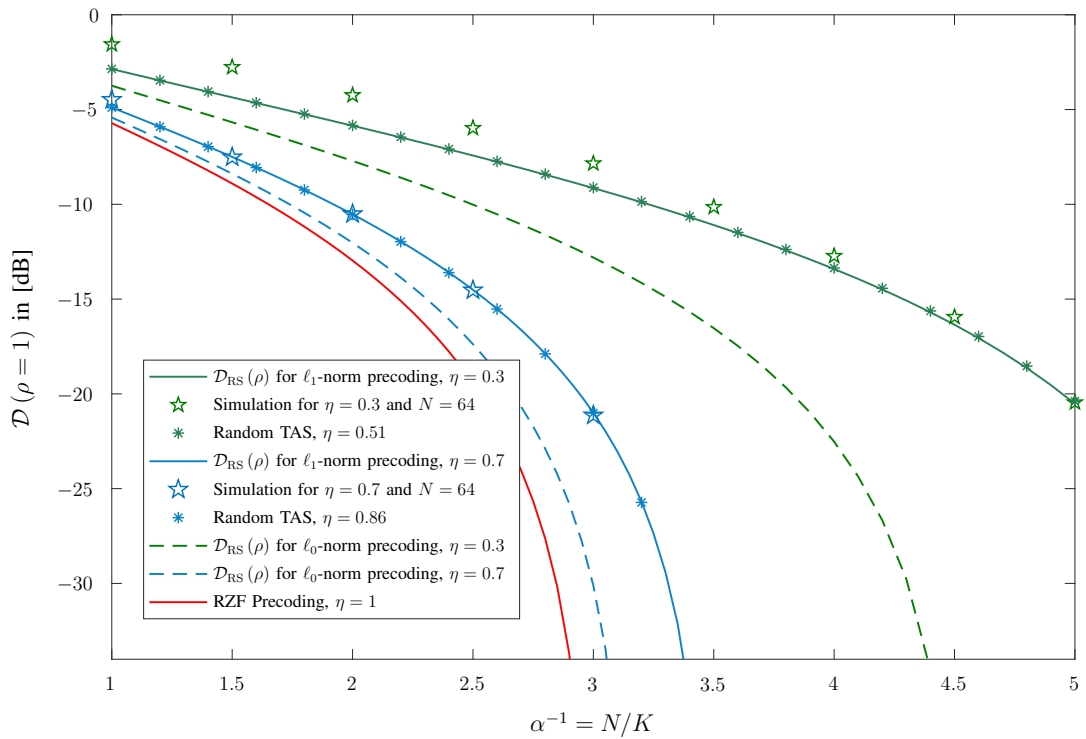}{
\psfrag{D in dB}[c][c][0.3]{$\sfD\prant{\rho=1}$ in [dB]}
\psfrag{alpha-inv}[c][c][0.3]{$\alpha^{-1}=N/K$}
\psfrag{RRREEE030DDDDSSSSBBBLO}[l][l][0.22]{$\sfD_{\rm RS} \prant{\rho}$ for $\ell_1$-norm precoding, $\eta=0.3$}
\psfrag{RRRSSS030DDDDSSSSBBBAA}[l][l][0.22]{Simulation for $\eta=0.3$ and $N=64$}
\psfrag{RRRNNN051DDDDSSSSBBBAA}[l][l][0.22]{Random \ac{tas}, $\eta=0.51$}
\psfrag{RRREEE070DDDDSSSSBBBLO}[l][l][0.22]{$\sfD_{\rm RS} \prant{\rho}$ for $\ell_1$-norm precoding, $\eta=0.7$}
\psfrag{RRRSSS070DDDDSSSSBBBAA}[l][l][0.22]{Simulation for $\eta=0.7$ and $N=64$}
\psfrag{RRRNNN086DDDDSSSSBBBAA}[l][l][0.22]{Random \ac{tas}, $\eta=0.86$}
\psfrag{RRREEE030DDDDSSSSBBBBA}[l][l][0.22]{$\sfD_{\rm RS} \prant{\rho}$ for $\ell_0$-norm precoding, $\eta=0.3$}
\psfrag{RRREEE070DDDDSSSSBBBBA}[l][l][0.22]{$\sfD_{\rm RS} \prant{\rho}$ for $\ell_0$-norm precoding, $\eta=0.7$}
\psfrag{RRREEE100DDDDSSSSBBBBA}[l][l][0.22]{RZF Precoding, $\eta=1$}
\psfrag{-5}[r][c][0.25]{$-5$}
\psfrag{-10}[r][c][0.25]{$-10$}
\psfrag{-15}[r][c][0.25]{$-15$}
\psfrag{-20}[r][c][0.25]{$-20$}
\psfrag{-25}[r][c][0.25]{$-25$}
\psfrag{-30}[r][c][0.25]{$-30$}
\psfrag{0}[r][c][0.25]{$0$}
%
\psfrag{1}[c][b][0.25]{$1$}
\psfrag{1.5}[c][b][0.25]{$1.5$}
\psfrag{1.4}[c][b][0.25]{$1.4$}
\psfrag{1.6}[c][b][0.25]{$1.6$}
\psfrag{5}[c][b][0.25]{$5$}
\psfrag{2}[c][b][0.25]{$2$}
\psfrag{2.5}[c][b][0.25]{$2.5$}
\psfrag{3.5}[c][b][0.25]{$3.5$}
\psfrag{4}[c][b][0.25]{$4$}
\psfrag{4.5}[c][b][0.25]{$4.5$}
\psfrag{3}[c][b][0.25]{$3$}

}}
\caption{Asymptotic distortion versus $\alpha^{-1}=N/K$ for the \ac{glse} precoder with $\ell_1$-norm penalty when the average power is $\sfp=0.5$. The simulations for $N=64$ shows that the \ac{rs} solution consistently tracks the performance of finite dimensional~systems.}
\label{fig:L1_Dist}
\end{figure}

The asymptotic distortion for this precoder has been plotted as a function of the inverse load in Fig.~\ref{fig:L1_Dist}. For sake of comparison, we have also sketched the curve for the case with $\ell_0$-norm penalty, as well as random \ac{tas}. As it is observed, the computationally feasible \ac{glse} precoder based on $\ell_1$-norm minimization consistently tracks the optimal performance with a slight degradation. To get a quantitative comparison, we have fitted the distortion curve with the one given for random \ac{tas}. The performance of \ac{glse} precoding with $\ell_1$-norm penalty at $\eta=0.3$ is tracked via random \ac{tas} with $\eta=0.51$. This means that by using LASSO regression $0.21N$ less active antennas are required compared to random \ac{tas}. Thus, $0.14N$ less antennas are saved than with the optimal selection algorithm. Nevertheless, we still get a significant enhancement compared to the benchmark performance.

In order to validate the result given via the replica method, we have further determined the distortion of the precoder numerically using CVX \cite{cvx,gb08}. The simulations are given for the same set of system parameters considering $N=64$ transmit antennas and show that the asymptotic results accurately match the performance of the precoder even in finite dimensions. The values of $\lambda$ and $\lambda_1$ in the simulations are tuned for the given power and \ac{tas} constraints via the tuning strategy illustrated in \ref{sec:ex1}. In this strategy, we find $\lambda$ and $\lambda_1$ such that the asymptotic average transmit power and the fraction of active antennas given by the replica method satisfy the given constraints. The accurate consistency of the results, therefore, validates our tuning strategy and indicates that the large-system results can be employed to tune \ac{glse} precoders in finite dimensions.

\begin{figure}[t]
\centering
\resizebox{1\linewidth}{!}{
\pstool[width=.35\linewidth]{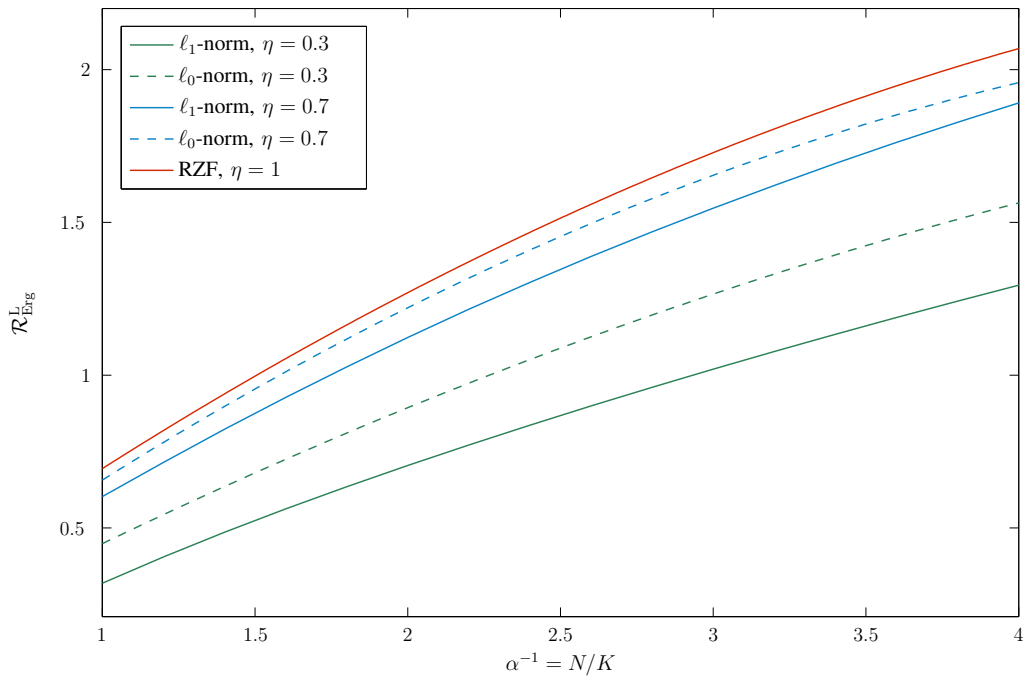}{
\psfrag{MI}[c][c][0.25]{$\mar_\av^{\rm L}$}
\psfrag{alpha-inv}[c][c][0.25]{$\alpha^{-1}=N/K$}
\psfrag{LOLNNNE030MMM}[l][l][0.25]{$\ell_1$-norm, $\eta=0.3$}
\psfrag{ZZZNNNE030MMM}[l][l][0.25]{$\ell_0$-norm, $\eta=0.3$}
\psfrag{LOLNNNE070MMM}[l][l][0.25]{$\ell_1$-norm, $\eta=0.7$}
\psfrag{ZZZNNNE070MMM}[l][l][0.25]{$\ell_0$-norm, $\eta=0.7$}
\psfrag{ZZZNNNE100MMM}[l][l][0.25]{\ac{rzf}, $\eta=1$}


\psfrag{0.5x}[r][c][0.23]{$0.5$}
\psfrag{1x}[r][c][0.23]{$1$}
\psfrag{1.5x}[r][c][0.23]{$1.5$}
\psfrag{2x}[r][c][0.23]{$2$}

%
\psfrag{1}[c][b][0.23]{$1$}
\psfrag{1.5}[c][b][0.23]{$1.5$}
\psfrag{1.4}[c][b][0.23]{$1.4$}
\psfrag{1.6}[c][b][0.23]{$1.6$}
\psfrag{1.8}[c][b][0.23]{$1.8$}
\psfrag{2}[c][b][0.23]{$2$}
\psfrag{2.5}[c][b][0.23]{$2.5$}
\psfrag{3.5}[c][b][0.23]{$3.5$}
\psfrag{2.6}[c][b][0.23]{$2.6$}
\psfrag{4}[c][b][0.23]{$4$}
\psfrag{3}[c][b][0.23]{$3$}

}}
\caption{Lower bound on the achievable ergodic rate per user versus $\alpha^{-1}=N/K$ for the \ac{glse} precoders with $\ell_0$- and $\ell_1$-norm penalty when the average transmit power is $\sfp=0.5$. $\mar_\av^{\rm L}$ has been maximized over all power control factors $\rho$ numerically. The average \ac{snr} is defined as $\snr \coloneqq \sfp/\sigma^2$ and is set to be $\log \snr=0$ dB.}
\label{fig:rate_L0L1}
\end{figure}

From Lemma~\ref{lemma1}, one can calculate the lower bound on the achievable ergodic rate of the \ac{glse} precoders given in Sections~\ref{sec:ex1} and \ref{sec:ex2}. This lower bound has been sketched versus the inverse load in Fig.~\ref{fig:rate_L0L1}. Here, the average \ac{snr} is defined as $\snr\coloneqq \sfp/\sigma^2$ and is set to $\log\snr=0$ dB. The lower bound has been optimized in terms of $\rho$. 

\subsection{Antenna Selection with Restricted \ac{papr}}
\label{sec:ex3}
In practice, the \ac{rf}-chains are restricted in terms of the peak transmit power. This restriction comes from the fact that the efficiency of power amplifiers significantly reduces for large output back-offs\footnote{For a power amplifier, the output back-off is defined as the maximum possible output power divided by the mean power at which the amplifier performs linearly.}. It is therefore desired in practice to transmit signals with limited \ac{papr} over the active antennas. In this respect, considering the \ac{glse} precoders in the previous sections, the assumption of $\setX=\setC$ is an inaccurate model for many systems. The inaccuracy turns to be more pivotal when the transmit signal is desired to have a relatively small \ac{papr}. To address this issue, one can modify the~\ac{glse} precoders in Sections \ref{sec:ex1} and \ref{sec:ex2} by setting $\setX$ to be
\begin{align}
\setX = \set{ x: \hspace*{2mm}   \abs{x} \leq \sqrt{P}}. \label{eq:SetX_PAPR}
\end{align}
Here, the output symbols are restricted to lie inside a circle with radius $\sqrt{P}$ which means that the per-antenna peak transmit power is upper bounded by $P$. Consequently, one can address both \ac{tas} and the \ac{papr} restriction by considering a penalty function as in \eqref{eq:penalty0} or \eqref{eq:penalty_ell_1} and letting the precoding support be as in \eqref{eq:SetX_PAPR}. We study these forms of the \ac{glse} precoders~in~the~following.

\subsubsection*{Optimal \ac{tas} for Restricted \ac{papr}}
For optimal \ac{tas}, the penalty is set to \eqref{eq:penalty0}. In this case, the decoupled precoder is given by
\begin{align}
\glse_\dec(s^\rs|\xi)=
\begin{cases}
  \dfrac{ s^\rs}{\abs{s^\rs}} \sqrt{P} \qquad  & \hat{\tau}_0 \leq \abs{s^\rs} \\
    0             & \tilde{\tau}_0 \leq \abs{s^\rs} < \hat{\tau}_0 \\
    \dfrac{s^\rs}{1+\xi\lambda} &\tau_0 \leq \abs{s^\rs}\leq \tilde{\tau}_0 \\
    0             &0\phantom{_0}\leq \abs{s^\rs} < \tau_0 \label{eq:single1}
\end{cases}
\end{align}
where $s^\rs\sim \mathcal{CN}(0, {\rho^\rs})$ with $\rho^\rs$ in \eqref{eq:rho_RS_AA} and the thresholds $\tau_0$, $\tilde{\tau}_0$ and $\hat{\tau}_0$ read
\begin{subequations}
\begin{align}
\tau_0&\coloneqq\sqrt{\xi \lambda_0 (1+\xi\lambda)} \label{eq:tau} \\
\tilde{\tau}_0&\coloneqq (1+\xi\lambda)\sqrt{P} \label{eq:tau_t} \\
\hat{\tau}_0&\coloneqq\max \set{ (1+\xi\lambda)\sqrt{P},\frac{1+\xi\lambda}{2} \sqrt{P} +\frac{\xi\lambda_0}{2\sqrt{P}}} \label{eq:tau_h}
\end{align}
\end{subequations}
for $\xi\coloneqq\alpha^{-1}(1 + \chi)$. Consequently, the average transmit power is determined as
\begin{align}
\sfp = \frac{\rho^\rs+\tau_0^2}{(1+\xi\lambda)^2} \exp\set{-\frac{\tau_0^2}{\rho^\rs}}
-\frac{\rho^\rs+\tilde{\tau_0}^2}{(1+\xi\lambda)^2} \exp\set{-\frac{\tilde{\tau}_0^2}{\rho^\rs}}
 + P \hspace*{.5mm} \exp\set{-\frac{\hat{\tau}_0^2}{\rho^\rs}}, \label{eq:p_1}
\end{align}
and the asymptotic fraction of active transmit antennas reads
\begin{align}
\sfa =\exp\set{-\frac{\tau_0^2}{\rho^\rs}}+\exp\set{-\frac{\hat{\tau}_0^2}{\rho^\rs}}-\exp\set{-\frac{\tilde{\tau}_0^2}{\rho^\rs}}. \label{eq:a_1}
\end{align}
The scalar $\chi$ further satisfies
\begin{align}
\hspace*{-2mm} \chi= \frac{\xi}{\rho^\rs} \frac{\rho^\rs \hspace*{-.5mm} + \hspace*{-.5mm} \tau_0^2}{1\hspace*{-.5mm}+ \hspace*{-.5mm}\xi\lambda} \left[ \exp\set{-\frac{\tau_0^2}{\rho^\rs}}
\hspace*{-.5mm} - \hspace*{-.5mm} \exp\set{-\frac{\tilde{\tau}_0^2}{\rho^\rs}} \right] \hspace*{-.7mm} + \hspace*{-.7mm} \frac{\xi\hat{\tau}_0\sqrt{P}}{\rho^\rs}  \hspace*{-.7mm} \exp\set{-\frac{\hat{\tau}_0^2}{\rho^\rs}} \hspace*{-.7mm} + \hspace*{-.7mm} \xi \sqrt{\frac{\pi P}{\rho^\rs}} \rmQ(\sqrt{\frac{2}{\rho^\rs}}\hat{\tau}_0). \label{eq:chi_1}
\end{align}

Comparing the decoupled precoder to \eqref{eq:sing0}, the decoupled transmit symbol in this case is obtained from $s^\rs$ by two steps of thresholding. In the first step, the decoupled input $s^\rs$ is compared to the threshold $\tilde{\tau}_0$, in order to be constrained \ac{wrt} the peak power $P$. The second step, then, imposes the \ac{tas} constraint as in \eqref{eq:sing0} using the thresholds $\tau_0$ and $\hat{\tau}_0$. Consequently, $\tau_0$ and $\hat{\tau}_0$ depend on the selection factor $\lambda_0$ while the threshold $\tilde{\tau}_0$ is only controlled by the peak power $P$. As $\lambda_0$ tends to zero, we have $\tau_0 \downarrow 0$ and $\hat{\tau}_0 = \tilde{\tau}_0$, and thus, the precoder reduces to the \ac{papr} limited precoder studied in \cite{sedaghat2017lse}. By fixing some constraints on the fraction of active transmit antennas and the \ac{papr}, $\lambda$ and $\lambda_0$ are tuned as discussed in Section \ref{sec:ex1}. 

\subsubsection*{\ac{tas} with $\ell_1$-norm Minimization for Restricted \ac{papr}}
For the precoding support in \eqref{eq:SetX_PAPR},~the optimization problem in \ac{glse} precoding with penalty \eqref{eq:penalty_ell_1} is convex for a large range of constraints on the average power and the fraction of active antennas. With this penalty,~the~decoupled precoder is derived as
\begin{align}
\glse_\dec(s^\rs|\xi)=
\begin{cases}
 \dfrac{s^\rs}{\abs{s^\rs}} \sqrt{P} &\tilde{\tau}_1\leq \abs{s^\rs}\\
    \dfrac{s^\rs}{1+\xi\lambda} \dfrac{\abs{s^\rs}-\tau_1}{\abs{s^\rs}}   &\tau_1 \leq \abs{s^\rs} < \tilde{\tau}_1 \\
    0             &\hspace*{.8mm}0\hspace*{.3mm}\leq \abs{s^\rs} < \tau_1 \label{eq:single_norm11}
\end{cases}
\end{align}
where $\xi\coloneqq\alpha^{-1}(1 + \chi)$. Similar to the case with $\ell_0$-norm penalty, the decoupled precoder constructs the decoupled transmit symbol by thresholding the input $s^\rs\sim \mathcal{CN}(0, {\rho^\rs})$ in two steps. In the first step, the magnitude of the decoupled input is compared to 
\begin{align}
\tilde{\tau}_1 \coloneqq \sqrt{P} \prant{ 1+\xi\lambda}+{\xi \lambda_1}/{2}
\end{align}
such that the peak power is limited to $P$. Then, a soft thresholding is performed \ac{wrt} $\tau_1\coloneqq {\xi \lambda_1}/{2}$ for \ac{tas}. The average transmit power of the precoder in this case is 
\begin{align}
\sfp=\dfrac{\rho^\rs}{\left(1+\xi\lambda \right)^2} \left[ \exp\set{-\frac{\tau_1^2}{\rho^\rs}} - \exp\set{-\frac{\tilde{\tau}_1^2}{\rho^\rs}} \right] -2 \dfrac{\tau_1\sqrt{\pi \rho^\rs}}{\left(1+\xi\lambda \right)^2} \left[ \rmQ ( \sqrt{\frac{2}{\rho^\rs}} \tau_1 ) - \rmQ ( \sqrt{\frac{2}{\rho^\rs}} \tilde{\tau}_1 ) \right]
\end{align}
and the asymptotic fraction of active transmit antennas is calculated by
\begin{align}
\eta = \exp\set{-\frac{\tau_1^2}{\rho^\rs}}.
\end{align}
For this case, the scalar $\chi$ fulfills the fixed-point equation
\begin{align}
\frac{\alpha \chi}{1+\chi} \left(1+\xi\lambda\right)= \exp\set{-\frac{\tau_1^2}{\rho^\rs}} \hspace*{-.5mm} - \hspace*{-.5mm} \exp\set{-\frac{\tilde{\tau}_1^2}{\rho^\rs}} \hspace*{-.5mm} - \hspace*{-.5mm} \dfrac{\sqrt{\pi \rho^\rs}}{\rho^\rs } \left[ \tau_1 \rmQ ( \sqrt{\frac{2}{\rho^\rs}} \tau_1 ) \hspace*{-.5mm} - \hspace*{-.5mm} \tilde{\tau}_1 \rmQ ( \sqrt{\frac{2}{\rho^\rs}} \tilde{\tau}_1 )\right].
\end{align}
As in the case with optimal \ac{tas}, for $\lambda_1 \downarrow 0$ the threshold $\tau_1$ tends to zero and the decoupled precoder describes the single-user representation of the \ac{papr}-limited precoder. 

\subsubsection*{Numerical investigations}
\begin{figure}[t]
\hspace*{-1.3cm}  
\centering
\resizebox{1.05\linewidth}{!}{
\pstool[width=.35\linewidth]{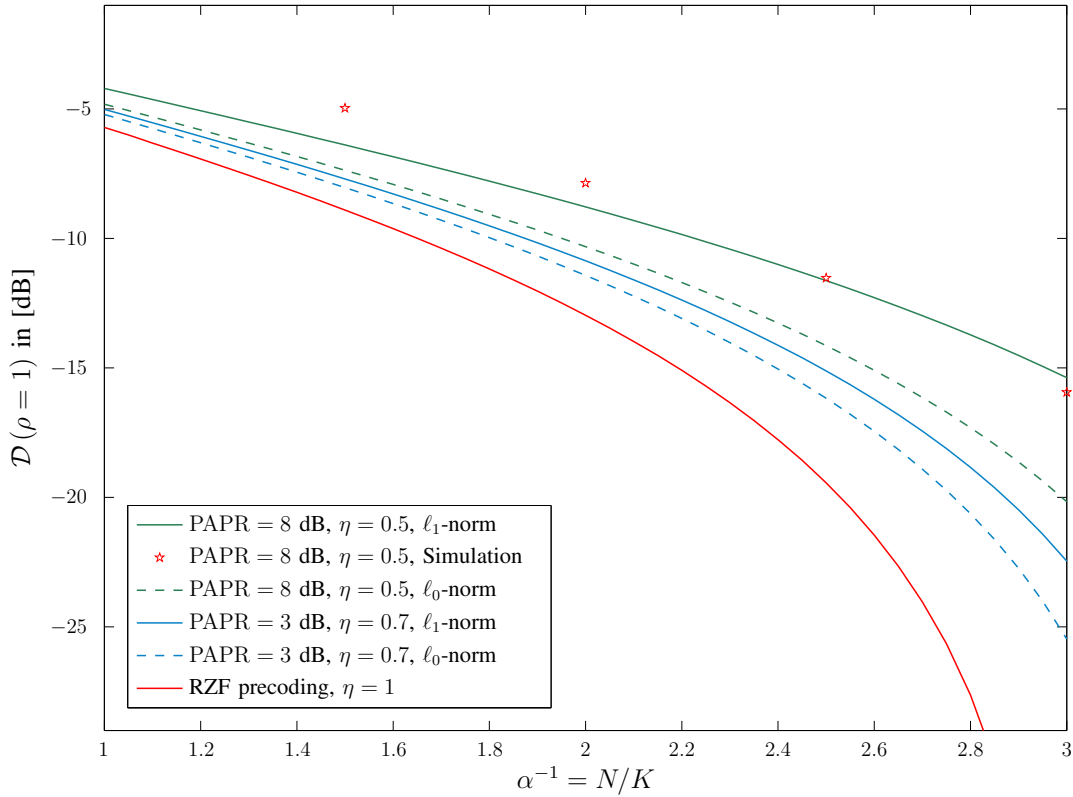}{
\psfrag{D in dB}[c][c][0.3]{$\sfD\prant{\rho=1}$ in [dB]}
\psfrag{alpha-inv}[c][c][0.3]{$\alpha^{-1}=N/K$}
\psfrag{PAPR=8dB-eta=0.5-L1NormXXXAA}[l][l][0.25]{$\papr=8$ dB, $\eta=0.5$, $\ell_1$-norm}
\psfrag{PAPR=8dB-eta=0.5-L1SImuXXXAA}[l][l][0.25]{$\papr=8$ dB, $\eta=0.5$, Simulation}
\psfrag{PAPR=8dB-eta=0.5-Z0NormXXXAA}[l][l][0.25]{$\papr=8$ dB, $\eta=0.5$, $\ell_0$-norm}
\psfrag{PAPR=3dB-eta=0.7-L1NormXXXAA}[l][l][0.25]{$\papr=3$ dB, $\eta=0.7$, $\ell_1$-norm}
\psfrag{PAPR=3dB-eta=0.7-Z0NormXXXAA}[l][l][0.25]{$\papr=3$ dB, $\eta=0.7$, $\ell_0$-norm}
\psfrag{PAPR=RZF-eta=1.0-Z0NormXXXAA}[l][l][0.25]{RZF precoding, $\eta=1$}
\psfrag{PAPR=0dB-eta=0.5-Z0NormXXXAA}[l][l][0.25]{$\papr=0$ dB, $\eta=0.5$, $\ell_0$-norm}
\psfrag{-5}[r][c][0.23]{$-5$}
\psfrag{-10}[r][c][0.23]{$-10$}
\psfrag{-15}[r][c][0.23]{$-15$}
\psfrag{-20}[r][c][0.23]{$-20$}
\psfrag{-25}[r][c][0.23]{$-25$}
\psfrag{-30}[r][c][0.23]{$-30$}
\psfrag{0}[r][c][0.23]{$0$}
%
\psfrag{1}[c][b][0.23]{$1$}
\psfrag{1.2}[c][b][0.23]{$1.2$}
\psfrag{1.4}[c][b][0.23]{$1.4$}
\psfrag{1.6}[c][b][0.23]{$1.6$}
\psfrag{1.8}[c][b][0.23]{$1.8$}
\psfrag{2}[c][b][0.23]{$2$}
\psfrag{2.2}[c][b][0.23]{$2.2$}
\psfrag{2.4}[c][b][0.23]{$2.4$}
\psfrag{2.6}[c][b][0.23]{$2.6$}
\psfrag{2.8}[c][b][0.23]{$2.8$}
\psfrag{3}[c][b][0.23]{$3$}

}}
\caption{Asymptotic distortion versus the inverse load $\alpha^{-1}=N/K$ for the \ac{glse} precoder with restricted \ac{papr} when the average transmit power is set to $\sfp=0.5$. The solid asnd dashed lines denote the \ac{rs} solution which track closely the exact performance. The simulations are given for $N=64$ in which the precoder is tuned using the replica solution.} 
\label{fig:PAPR-Dist}
\end{figure}

Fig.~\ref{fig:PAPR-Dist} shows the asymptotic distortion of \ac{papr}-limited \ac{glse} precoders with $\ell_0$-norm and $\ell_1$-norm penalties in terms of the inverse load $\alpha^{-1}=N/K$. The average transmit power is set to $\sfp=0.5$, and the peak power is chosen such that the \ac{papr} constraint is fulfilled. Moreover, the power control factor is fixed to ${\rho=1}$ and the \ac{papr} has been determined as $\papr=\log\left(P/\sfp \right)$. The curves for \ac{papr}-limited precoders with random \ac{tas} have been furthermore fitted numerically to the \ac{glse} precoders. 

It is observed that at higher \ac{papr}s, e.g. $\papr=8$ dB, the precoder outperforms random \ac{tas} with approximately the same reduction in the fraction of active antennas as in cases with no \ac{papr} constraint. However, for smaller \ac{papr}s, this advantage reduces. It is moreover observed that for small \ac{papr}s the degradation caused by replacing $\ell_0$-norm with $\ell_1$-norm is decreased. For example, as Fig.~\ref{fig:PAPR-Dist} depicts, at $\papr=3$~dB the reduction in the number of active antennas compared to random \ac{tas} is around $0.25N$ for $\ell_0$-norm and $0.2N$ for $\ell_1$-norm penalty. 



\subsection{Antenna Selection for Discrete Constellations}
\label{sec:MPSK}
Some new suggestions for \ac{mimo} transmitters have proposed structures whose transmit signal is taken from a discrete constellation. An example is the \ac{lmsrf} transmitter in which multiple load modulators, fed by a single \ac{rf}-chain, construct the transmit constellation \cite{sedaghat2014novel,sedaghat2016discrete,sedaghat2016load}. Each load-modulator in this case is equipped with some switches, and the cardinality of the transmit constellation is restricted by the total number of transmit states, e.g., for three switches, there are eight possible states. For these transmitters, conventional precoders cannot be employed, as the majority of the schemes in the literature sets $\setX=\setC$ and bounds the transmit amplitude by further processing. Another example is a \ac{mimo} system with low-resolution digital-to-analog converters where the precoding support is restricted with the output of the digital-to-analog converters \cite{jacobsson2017quantized}.

The generality of the precoding  support in the GLSE scheme, however, enables us to precode the data directly over the constellations of these transmitters as well. In \cite{sedaghat2017lse}, a special class of \ac{glse} precoders, i.e., nonlinear \ac{lse} precoders, was considered to address the $M$-\ac{psk} constellations when the whole transmit antennas are set active. We extend the analysis to the case with \ac{tas}. To this end, we set 
\begin{align}
\setX = \set{ 0} \cup \set{ \sqrt{P} \hspace{.2mm} \exp\set{\rmj\frac{2k \pi}{M}} \hspace*{2mm} \text{for} \hspace*{2mm} k\in [1:M] }. \label{eq:XMpsk}
\end{align}
The precoder in this case maps the data to a vector whose symbols are either taken from an $M$-\ac{psk} constellation or are zero. In general, the number of active transmit antennas is restricted by the penalty in \eqref{eq:penalty0}. Nevertheless, for the $M$-\ac{psk} constellation, $\ell_2$- and $\ell_0$-norm are related via $\norm{\bx}^2=P \norm{\bx}_0$ indicating that any restriction on the transmit power limits the number of active antennas. Consequently, \ac{tas} can be enforced in this case by the penalty function\footnote{Note that for some other discrete constellations, this simplification may not work and one needs to keep the penalty as in \eqref{eq:penalty0}, e.g., for $16$-QAM.}
\begin{align}
u(\bv)=\lambda\norm{\bv}^2. \label{eq:uMpsk}
\end{align}
For this \ac{glse} precoder, the decoupled precoder is given by
\begin{align}
\glse_\dec(s^\rs|\xi)=
\begin{cases}
     \sqrt{P} \hspace{.2mm} \exp\set{\rmj\tfrac{2k^\star \pi}{M}} &\abs{s^\rs}\geq \tau_\sfd \\
    0             &\abs{s^\rs} < \tau_\sfd \label{eq:single_psk}
\end{cases}
\end{align}
with $s^\rs\sim \mathcal{CN}(0, {\rho^\rs})$ and $\rho^\rs$ given in \eqref{eq:rho_RS_AA}, where the threshold $\tau_\sfd$ is defined as
\begin{align}
\tau_\sfd\coloneqq \frac{\sqrt{P} \prant{1+\xi\lambda} }{2 \hspace*{.3mm} \Theta \prant{ k^\star| \sphericalangle s^\rs}} 
\end{align}
for $\xi\coloneqq\alpha^{-1}(1 + \chi)$ and $k^\star \coloneqq \arg\max_{k} \Theta \prant{ k| \sphericalangle s^\rs}$ with
\begin{align}
\Theta\prant{k|\theta} \coloneqq \cos\left( \frac{2k\pi}{M}-\theta \right).
\end{align}
The decoupled precoder in this case describes a thresholding operator over the $M$-\ac{psk} constellation in which the magnitude of the decoupled input is first compared to $\tau_\sfd$, and then, its non-zero output is mapped to an $M$-\ac{psk} symbol whose phase is closest to $\sphericalangle s^\rs$. The asymptotic fraction of active antennas for this precoder is given by
\begin{align}
\eta = \frac{1}{2\pi} \int_{0}^{2\pi} \exp\set{ - \frac{P\prant{1+\xi \lambda}^2}{4\rho^\rs \left[\max_k \Theta(k|\theta)\right]^2 } } \dif \theta
\end{align}
and $\sfp=P\eta$. The fixed-point equation for $\chi$ moreover reads
\begin{align}
\hspace*{-3mm}\frac{\alpha \chi}{1+\chi} \rho^\rs = \frac{P\prant{1+\xi \lambda}}{2} \eta + \frac{1}{2\pi} \int_{0}^{2\pi} \sqrt{\pi \rho^\rs P} \left[\max_k \Theta(k|\theta)\right] \rmQ\prant{  \frac{\sqrt{P}\prant{1+\xi \lambda}}{\sqrt{2\rho^\rs} \left[\max_k \Theta(k|\theta)\right] } } \dif \theta.
\end{align}
In this case, by growth of $\lambda$, the threshold $\tau_d$ increases, and thus, the fraction of active transmit antennas reduces. The asymptotics of constant envelope transmission with \ac{tas} is moreover derived by taking the limit $M\uparrow\infty$. We determine this limit in Appendix~\ref{app:entropy}.~For~a~given constraint on $\eta$, the factor $\lambda$ is tuned via the tuning strategy illustrated in~Section~\ref{sec:ex1}.

\subsubsection*{Numerical investigations}
For BPSK signals, i.e. $M=2$, the decoupled precoder reduces to 
\begin{align}
\glse_\dec(s^\rs|\xi)=\sqrt{P} \hspace{.2mm} \exp\set{\rmj\tfrac{2k^\star \pi}{M}} T (\re{s^\rs}|\xi)
\end{align}
where the real-valued thresholding function $T (\cdot|\xi)$ is defined as
\begin{align}
T (u|\xi)=
\begin{cases}
    1     		 &\qquad \abs{u} \geq  \sqrt{P}\prant{1+\xi \lambda}/2 \\
    0            &\qquad \abs{u} <  \sqrt{P}\prant{1+\xi \lambda}/2 \label{eq:single_psk}
\end{cases}.
\end{align}
Considering the QPSK constellation, i.e. $M=4$, the decoupled \ac{glse} precoder reads 
\begin{align}
\glse_\dec(s^\rs|\xi)=\sqrt{P} \hspace{.2mm} \exp\set{\rmj\tfrac{2k^\star \pi}{M}} T (\re{s^\rs}|\xi) T (\img{s^\rs}|\xi)
\end{align}

\begin{figure}[t]
\centering
\resizebox{.94\linewidth}{!}{
\pstool[width=.35\linewidth]{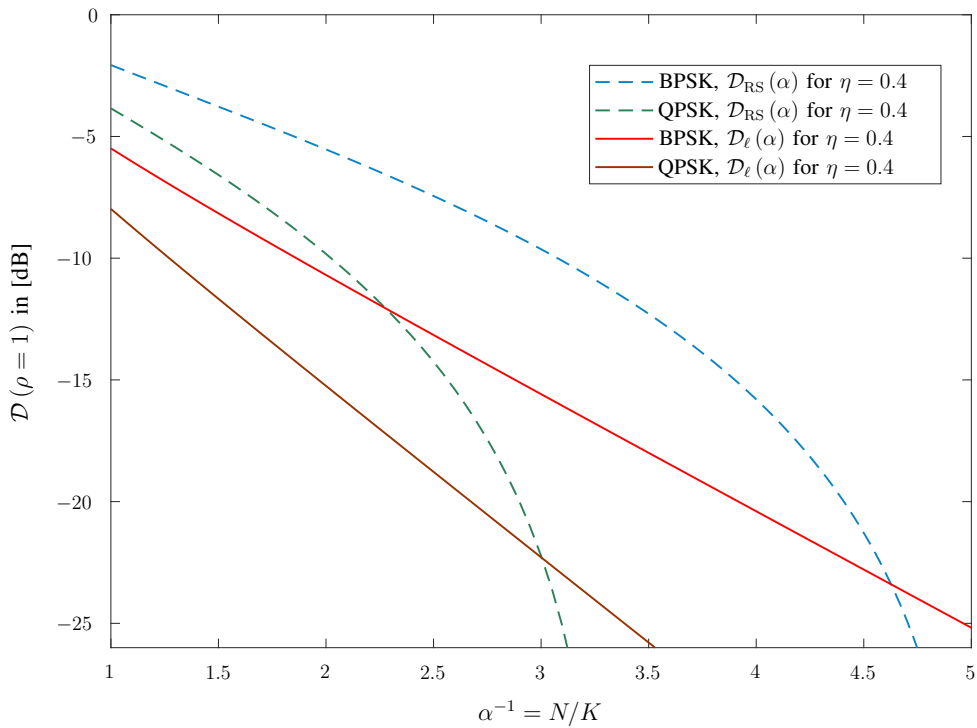}{
\psfrag{D in dB}[c][c][0.3]{$\sfD\prant{\rho=1}$ in [dB]}
\psfrag{alpha-inv}[c][c][0.3]{$\alpha^{-1}=N/K$}
\psfrag{BPSK-eta=0.4XXXA1BBBBBB}[l][l][0.27]{BPSK, $\sfD_{\rm RS} \left( \alpha \right)$ for $\eta=0.4$}
\psfrag{BPSK-LOWERXXXABBBBBB}[l][l][0.27]{BPSK, $\sfD_{\ell} \left( \alpha \right)$ for $\eta=0.4$}
\psfrag{QPSK-eta=0.4XXXA1BBBBBB}[l][l][0.27]{QPSK, $\sfD_{\rm RS} \left( \alpha \right)$ for $\eta=0.4$}
\psfrag{QPSK-LOWERXXXABBBBBB}[l][l][0.27]{QPSK, $\sfD_{\ell} \left( \alpha \right)$ for $\eta=0.4$}

\psfrag{-5}[r][c][0.25]{$-5$}
\psfrag{-10}[r][c][0.25]{$-10$}
\psfrag{-15}[r][c][0.25]{$-15$}
\psfrag{-20}[r][c][0.25]{$-20$}
\psfrag{-25}[r][c][0.25]{$-25$}
\psfrag{-30}[r][c][0.25]{$-30$}
\psfrag{0}[r][c][0.25]{$0$}
%
\psfrag{1}[c][b][0.25]{$1$}
\psfrag{1.5}[c][b][0.25]{$1.5$}
\psfrag{4.5}[c][b][0.25]{$4.5$}
\psfrag{5}[c][b][0.25]{$5$}
\psfrag{1.8}[c][b][0.25]{$1.8$}
\psfrag{2}[c][b][0.25]{$2$}
\psfrag{2.5}[c][b][0.25]{$2.5$}
\psfrag{2.4}[c][b][0.25]{$2.4$}
\psfrag{3.5}[c][b][0.25]{$3.5$}
\psfrag{4}[c][b][0.25]{$4$}
\psfrag{3}[c][b][0.25]{$3$}

}}
\caption{The \ac{rs} solution for the asymptotic distortion of BPSK and QPSK constellations and the lower bound $\sfD_{\ell} \left( \alpha \right)$ determined from Lemma~\ref{lemma2} versus $\alpha^{-1}=N/K$. $P$ is set such that the average transmit power $\sfp=\eta P=1$.} 
\label{fig:MPSK-RS}
\end{figure}

For these particular examples, the asymptotic distortion given by the replica solutions have been plotted for $P=1$ in Fig.~\ref{fig:MPSK-RS} against the inverse load. To investigate the tightness of the \ac{rs} solution, we extend the approach in \cite[Appendix D]{sedaghat2017least} and derive a rigorous lower bound on the distortion of the $M$-PSK transmission with \ac{tas} in Lemma~\ref{lemma2}.
\begin{lemma}
\label{lemma2}
Consider the \ac{glse} precoder with the precoding support given in \eqref{eq:XMpsk} and the penalty function in \eqref{eq:uMpsk}. For a given load $\alpha$ and power control factor $\rho$, the asymptotic distortion of the precoder, when the asymptotic fraction of active transmit antennas is $\eta$, is bounded from below by $\sfD_{\ell}$ which satisfies the fixed-point equation
\begin{align}
\frac{\sfD_\ell}{\rho+\eta P}=1+\frac{\log (1+M) }{\alpha} + \log \frac{\sfD_\ell}{\rho+\eta P}.
\end{align}
\end{lemma}
\begin{prf}
The proof is given in Appendix~\ref{app:lower}.
\end{prf}

\begin{figure}[t]
\centering
\resizebox{.94\linewidth}{!}{
\pstool[width=.35\linewidth]{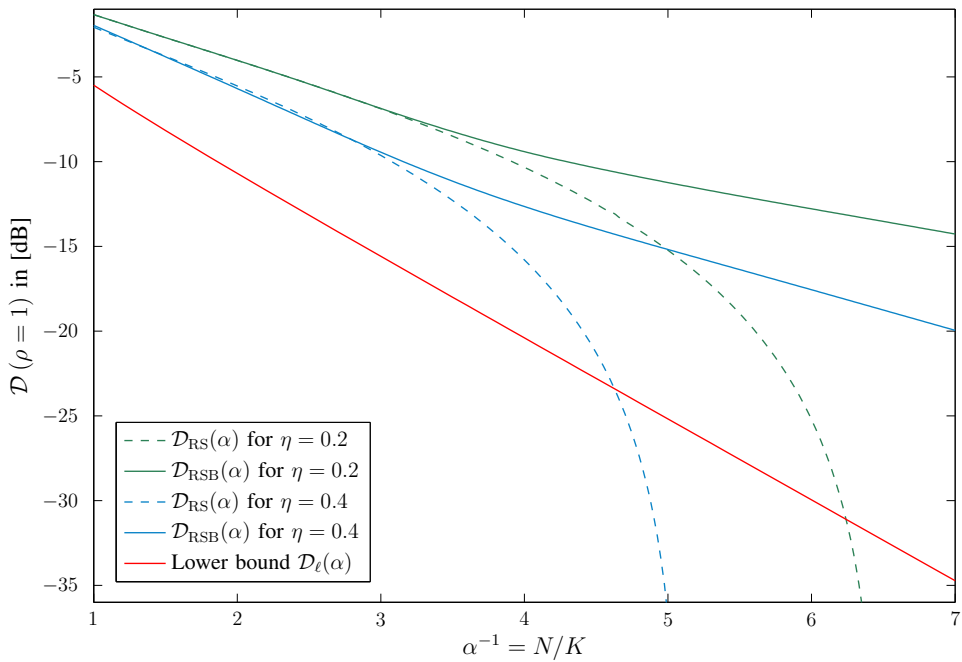}{
\psfrag{D in dB}[c][c][0.3]{$\sfD\prant{\rho=1}$ in [dB]}
\psfrag{alpha-inv}[c][c][0.3]{$\alpha^{-1}=N/K$}
\psfrag{BPSK-eta=0.4XXXA1}[l][l][0.27]{$\sfD_{\rm RS} (\alpha)$ for $\eta=0.4$}
\psfrag{BPSK-LOWERXXXA}[l][l][0.27]{Lower bound $\sfD_\ell (\alpha) $}
\psfrag{BPSK-eta=0.2XXXA1}[l][l][0.27]{$\sfD_{\rm RS} (\alpha)$ for $\eta=0.2$}
\psfrag{BPSK-eta=0.4XXRSB}[l][l][0.27]{$\sfD_{\rm RSB} (\alpha)$ for $\eta=0.4$}
\psfrag{BPSK-eta=0.2XXRSB}[l][l][0.27]{$\sfD_{\rm RSB} (\alpha)$ for $\eta=0.2$}

\psfrag{-5}[r][c][0.25]{$-5$}
\psfrag{-10}[r][c][0.25]{$-10$}
\psfrag{-15}[r][c][0.25]{$-15$}
\psfrag{-20}[r][c][0.25]{$-20$}
\psfrag{-25}[r][c][0.25]{$-25$}
\psfrag{-30}[r][c][0.25]{$-30$}
\psfrag{-35}[r][c][0.25]{$-35$}
\psfrag{0}[r][c][0.2]{$0$}
%
\psfrag{1}[c][b][0.25]{$1$}
\psfrag{5}[c][b][0.25]{$5$}
\psfrag{1.4}[c][b][0.25]{$1.4$}
\psfrag{6}[c][b][0.25]{$6$}
\psfrag{7}[c][b][0.25]{$7$}
\psfrag{2}[c][b][0.25]{$2$}
\psfrag{2.5}[c][b][0.25]{$2.5$}
\psfrag{2.4}[c][b][0.25]{$2.4$}
\psfrag{3.5}[c][b][0.25]{$3.5$}
\psfrag{4}[c][b][0.25]{$4$}
\psfrag{3}[c][b][0.25]{$3$}

}}
\caption{The one-step \ac{rsb} solution for the asymptotic distortion of BPSK constellation considering $\eta=0.2$ and $\eta=0.4$, as well as the lower bound $\sfD_{\ell}$ given by Lemma~\ref{lemma2}.}
\label{fig:BPSK-RSB}
\end{figure}


The lower bound given by Lemma~\ref{lemma2} clarifies the looseness of the \ac{rs} solution for moderate and large loads. In fact, as the figure shows the \ac{rs} solution outperforms the lower bound for small inverse loads. The solution, however, starts to violate the lower bound as the inverse load grows. Considering the discussions in Section~\ref{sec:discussion}, this observation indicates that the \ac{rs} assumption is not valid in this case, and therefore, one needs to study  \ac{rsb} solutions. We therefore have plotted the one-step \ac{rsb} solution in Fig.~\ref{fig:BPSK-RSB}. The figure demonstrates a larger range of inverse loads in which the replica solution outperforms the lower bound in Lemma~\ref{lemma2}. The one-step \ac{rsb} solution is determined by replacing the decoupled input with $s^{\rs} + s^{\rsb}_1$ and solving the fixed-point equations given in Proposition~\ref{thm:2}. Numerical investigations show that the one-step \ac{rsb} solution tracks simulations for a moderate load regime. It however starts to deviate as the inverse load increases; see discussions in \cite{sedaghat2017least}. Hence, for larger load regimes, more steps of \ac{rsb} are~required. 

\begin{figure}[t]
\centering
\resizebox{1\linewidth}{!}{
\pstool[width=.35\linewidth]{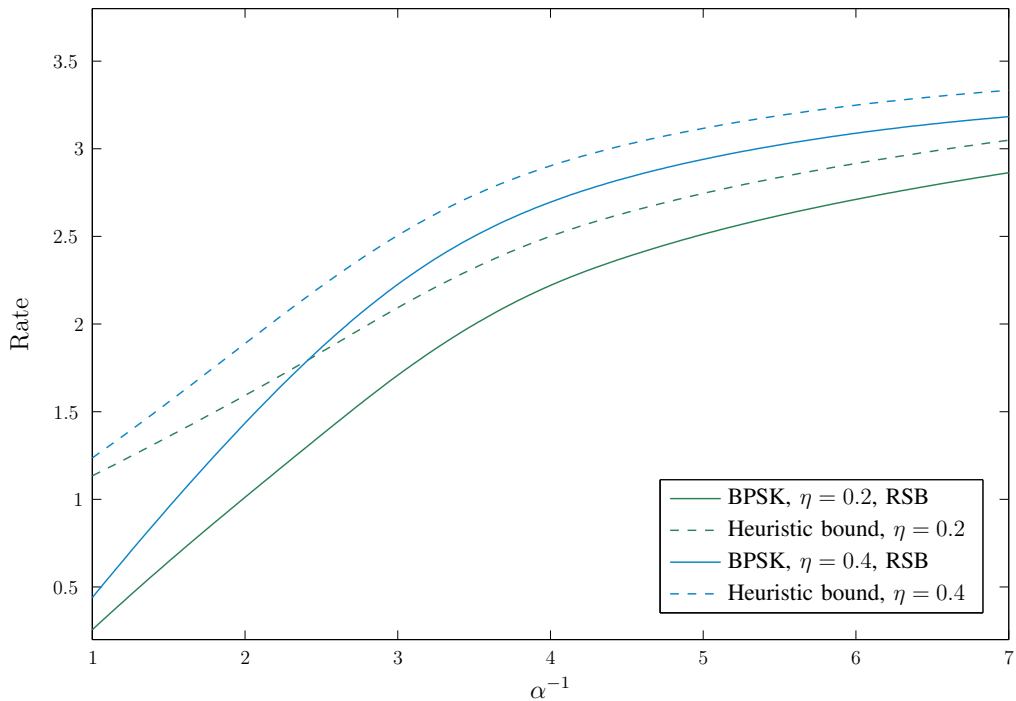}{
\psfrag{MI}[c][c][0.3]{$\rm Rate$}
\psfrag{alpha-inv}[c][c][0.3]{$\alpha^{-1}$}
\psfrag{UPXX04BBSSS-BPSK}[l][l][0.27]{Heuristic bound, $\eta=0.4$}
\psfrag{UPXX02BBSSS-BPSK}[l][l][0.27]{Heuristic bound, $\eta=0.2$}
\psfrag{UP02AABBSSS-BPSK}[l][l][0.27]{BPSK, $\eta=0.2$, RSB}
\psfrag{UP04AABBSSS-BPSK}[l][l][0.27]{BPSK, $\eta=0.4$, RSB}

\psfrag{0.5}[r][c][0.23]{$0.5$}
\psfrag{10}[r][c][0.23]{$-10$}
\psfrag{1.5}[r][c][0.23]{$1.5$}
\psfrag{-20}[r][c][0.23]{$-20$}
\psfrag{-25}[r][c][0.23]{$-25$}
\psfrag{2.5}[r][c][0.23]{$2.5$}
\psfrag{0}[r][c][0.23]{$0$}
%
\psfrag{1}[r][c][0.23]{$1$}
\psfrag{5}[c][b][0.23]{$5$}
\psfrag{1x}[c][b][0.23]{$1$}
\psfrag{6}[c][b][0.23]{$6$}
\psfrag{7}[c][b][0.23]{$7$}
\psfrag{2}[r][c][0.23]{$2$}
\psfrag{2x}[c][b][0.23]{$2$}
\psfrag{3x}[c][b][0.23]{$3$}
\psfrag{3.5}[r][c][0.23]{$3.5$}
\psfrag{4}[c][b][0.23]{$4$}
\psfrag{3}[r][c][0.23]{$3$}

}}
\caption{The lower bound on the ergodic rate considering the one-step \ac{rsb} solution. The \ac{snr} is set to $\log \snr= 10$ dB, $\rho=1$ and $P$ is chosen such that $\sfp=\eta P=1$. The dashed line shows the heuristic approximation of $\mar_\av$ assuming that the interference at the user terminals is independent of the data symbol.}
\label{fig:BPSK-RSB-MI}
\end{figure}

For the range of inverse loads over which the one-step \ac{rsb} solution outperforms $\sfD_\ell$ in Lemma~\ref{lemma2}, the lower bound on the achievable ergodic rate per user has been plotted assuming $\rho=1$ in Fig.~\ref{fig:BPSK-RSB-MI}. The noise power is considered to be $\sigma^2=0.1$. A heuristic bound on the achievable ergodic rate can be derived by assuming the interference caused by other users becoming independent of the data symbol in the large-system limit. In this case, one can approximate the ergodic rate when all the users are distributed uniformly in the cell as
\begin{align}
\mar_\av \approx \log\prant{1+\frac{\rho}{\sfp+\sfD(\rho)}}.
\end{align}
As Fig.~\ref{fig:BPSK-RSB-MI} depicts, $\mar_\av^{\rm L}$ lies close to this heuristic approximation as $N/K$ increases.

\section{Large-System Analysis}
\label{sec:large}
In this section, we state the derivations of the main results. To start with, consider the function
\begin{align}
\mae(\bv|\bs,\mH)=\norm{\mH \bv- \sqrt{\rho} \hspace*{.5mm}\bs}^2 +u(\bv), \label{eq:8}
\end{align}
which is referred to as the ``Hamiltonian''. We define the so-called partition function $\maz(\beta,h)$ as
\begin{align}
\maz(\beta,h) =\sum_{\bv} \exp\set{-\beta\mae(\bv|\bs,\mH)+hN\sfM^{\setW}_f(\bv;N)}
\end{align}
with $\sfM^{\setW}_f(\bv;N)$ being the marginal of the transmit signal given in Definition \ref{def:margin}. Moreover, we define the function $\maf(\beta,h)$ as
\begin{align}
\maf(\beta,h) \coloneqq \frac{1}{N}  \E \log \maz(\beta,h).
\end{align}
In the sequel, we show that both the asymptotic distortion and the asymptotic marginal of the transmit signal are directly derived from the function $\maf(\beta,h)$.
\subsection{Deriving the Asymptotic Distortion and Marginal}
\label{ssec:asy_marg}
We start by deriving the asymptotic marginal of transmit vector from $\maf(\beta,h)$. Our derivation is based on the following large deviations argument known as the saddle-point method \cite{dembo2010large}.
\begin{lemma}[Saddle-point method]\label{varahlemma}
Assume that $f(\bv): \setX^N \mapsto \setR$ is a bounded function, then
\begin{align}\label{Farhadkhan}
\min_{\bv \in \setX^N} f (\bv)
= - \lim_{\beta\uparrow \infty} \frac{1}{\beta} \log \int_{\setX^N} \exp\set{-\beta f( \bv )} \dif \bv
\end{align}
\end{lemma}
\begin{prf}
The lemma is concluded by  setting $\epsilon=1/\beta$, $\phi(x)=0$ and $\{\mu_{\epsilon}\}$ equal to a family of non-degenerate Gaussian measures with the rate function $f({x})$ in Varadhan's theorem \cite[Theorem~4.3.1]{dembo2010large}. The same result is concluded for discrete $\setX$ when we replace the integral~by~sum.
\end{prf}
Using Lemma~\ref{varahlemma}, it is straightforward to show that the asymptotic marginal reads
\begin{align}
\sfM^{\setW}_f(\bx) =\lim_{N\uparrow\infty}  \lim_{\beta\uparrow\infty}  \frac{\partial}{\partial h} \maf(\beta,h)|_{h=0}. \label{eq:10}
\end{align}
To describe the asymptotic distortion in terms of $\maf(\beta,h)$, we take expectation from both sides of \eqref{eq:8}. It is then concluded that the asymptotic distortion satisfies
\begin{align}
\alpha \sfD(\rho) + \sfM_u^\setT (\bx) = \tilde{\mae} \label{eq:13}
\end{align}
where $\sfM_u^\setT (\bx)$ denotes the asymptotic marginal of $u(\bx)$ over $\setT(n) \coloneqq [n]$, and $\tilde{\mae}$ is the average energy of the Hamiltonian in the large-system limit\footnote{In the context of statistical mechanics, $\tilde{\mae}$ is the average energy of the spin glass defined by the Hamiltonian \eqref{eq:8} in the thermodynamic limit.} defined as
\begin{align}
\tilde{\mae}={\lim_{N\uparrow\infty} \frac{1}{N} \E \mae(\bx|\bs,\mH)}.
\end{align}
From \eqref{eq:10}, $\sfM_u^\setT (\bx)$ is determined in terms of $\maf(\beta,h)$. Moreover, 
\begin{align}
\tilde{\mae} = - \lim_{n\uparrow\infty}  \lim_{\beta\uparrow\infty}  \frac{\partial}{\partial \beta} \maf(\beta,h)|_{h=0}. \label{eq:14}
\end{align}
Thus, the asymptotic distortion $\sfD(\rho)$ is evaluated from $\maf(\beta,h)$ considering the equality in \eqref{eq:13}.
\subsection{Analysis via the Replica Method}
Based on the discussions in Section \ref{ssec:asy_marg}, the large-system analysis of the \ac{glse} precoders reduces to determining $\maf(\beta,h)$. To this end, we employ the replica method which has been initially developed in statistical mechanics to study spin glasses \cite{edwards1975theory} and later employed in information theory to investigate the asymptotics of various problems; see \cite{tanaka2002statistical,guo2005randomly, rangan2012asymptotic, bereyhi2016rsb} and references therein. We start our analysis by noting that determining $\maf(\beta,h)$ needs the hard task of taking a logarithmic expectation to be overcome. The task is analytically non-tractable when the argument of the logarithm is a sum of exponential functions. Using the Riesz equality which states that for a non-negative random variable $\xx$ and real~$m$~\cite{riesz:30}
\begin{align}
\E \log \xx = \lim_{m\downarrow 0} \frac{1}{m} \log \E \xx^m,
\end{align}
and bypasses the logarithmic expectation by writing $\maf(\beta,h)$ as 
\begin{align}
\maf(\beta,h) = \frac{1}{N} \lim_{m\downarrow 0}  \frac{1}{m} \log \E \left[ \maz(\beta,h) \right]^m. \label{eq:18}
\end{align}
The computation of \eqref{eq:18} is still non-trivial, since the \ac{rhs} of \eqref{eq:18} needs to be determined for real values\footnote{More precisely, it should be determined at least for some real values of $m$ in a right neighborhood of zero} of $m$. The replica method determines \eqref{eq:18} utilizing the conjecture of the replica continuity. The replica continuity indicates that the analytic continuation of the non-negative integer moment function, i.e., the function 
\begin{align}
f_{\rm M}(m) \coloneqq \E \left[ \maz(\beta,h) \right]^m
\end{align}
with $m\in\setZ^+$, onto the set of non-negative real numbers equals to the non-negative real moment function, i.e., $f_{\rm M}(m)$ with domain $m\in\setR^+_0$. In other words, replica continuity suggests to determine the moment function for an integer $m$, and then, assume that the function is of the same form for $m\in\setR_0^+$. The rigorous justification of the replica continuity has not been yet precisely addressed in general; however, the analytic results from the theory of spin glasses confirm the validity of the conjecture for several cases. Considering the replica continuity, the moment function reads
\begin{subequations}
\begin{align}
f_{\rm M}(m)  &= \E \sum_{\set{\bv_a}} \prod_{a=1}^m \exp\set{-\beta\mae(\bv_a|\bs,\mH)+hN\sfM^{\setW}_f(\bv_a;N)} \\
&= \sum_{\set{\bv_a}}  \exp\set{hN\sum\limits_{a=1}^m \sfM^{\setW}_f(\bv_a;N)} \E \exp\set{-\beta\sum\limits_{a=1}^m \mae(\bv_a|\bs,\mH)}
\end{align}
\end{subequations}
where $\set{\bv_a}\coloneqq\set{\bv_1, \cdots, \bv_m}$ denotes the set of replicas\footnote{In fact, it is the reason that the method is called the ``replica'' method.}. By taking the expectation \ac{wrt} $\bs$, 
\begin{align}
f_{\rm M}(m) = \sum_{\set{\bv_a}} \E_{\mJ} \exp\set{-\beta\sum\limits_{a,b=1}^m \bv_a^\her \mJ \bv_b \xi_{ab} - N \Theta \set{\bv_a}} \label{eq:20}
\end{align}
where $\xi_{ab}\coloneqq \delta(a-b) - \rho \beta (1+\rho \beta m)^{-1}$, $\mJ$ denotes the Gramian of $\mH$, i.e. $\mJ\coloneqq \mH^\her \mH$, and
\begin{align}
\Theta \set{\bv_a} \coloneqq \frac{1}{N} \sum_{a=1}^m \left[ \beta u(\bv_a)- hN\sfM^{\setW}_f(\bv_a;N)\right] + \frac{1}{N}\Delta_m 
\end{align}
with $\Delta_m\coloneqq K \log (1+\rho \beta m)$. In order to take the expectation \ac{wrt} $\mJ$, we invoke the result reported in \cite{guionnet2005fourier} for spherical integrals, where a closed form formula has been given for some specific cases; see \cite[Appendix F]{bereyhi2016statistical} for more details on the spherical integrals. Let us consider the $m\times m$ ``replica correlation'' matrix $\mQ_m $ whose entries are defined as
\begin{align}
[\mQ_m]_{ab}=\frac{1}{N} \bv_a^\her \bv_b. \label{eq:Q}
\end{align}
Then, after taking the expectation \ac{wrt} $\mJ$, \eqref{eq:20} reduces to
\begin{align}
f_{\rm M}(m) = \sum_{\set{\bv_a}} \exp\set{-N \mg (\mT \mQ_m) - N \Theta \set{\bv_a}} \label{eq:23}
\end{align}
where $\mT \coloneqq \mI_m- \rho \beta (1+\rho \beta m)^{-1} \mone_m$, and $\mg(\cdot)$ reads
\begin{align}
\mg (\mM) &= \sum_{\ell=1}^m \int_0^{\beta \lambda_\ell} \rmR_{\mJ} (-\omega) \dif \omega \label{eq:24}
\end{align}
for some matrix $\mM_{m\times m}$ with eigenvalues $\set{\lambda_\ell}$ for $\ell\in[1:m]$. Here, $\rmR_\mJ (\cdot)$ is the $\rmR$-transform of the asymptotic eigenvalue distribution $\rmF_\mJ$. To determine the sum in \eqref{eq:23}, we divide the replicas $\set{\bv_a}$ into subshells \ac{wrt} their correlation matrices. More precisely, we define the subshell $\mas(\mQ)$ as the set of replicas $\set{\bv_a}$ whose correlation matrix is $\mQ$. In this case, \eqref{eq:23} reads
\begin{subequations}
\begin{align}
f_{\rm M}(m) &= \int \exp\set{-N \mg (\mT \mQ)} \left[ \sum_{\set{\bv_a} } \exp\set{ - N \Theta \set{\bv_a}} w(\mQ;\set{\bv_a}) \right] \dif \mQ \\
&= \int \exp\set{-N\mg (\mT \mQ)} \exp\set{- N\mai(\mQ)} \dif \mQ \label{eq:25}
\end{align}
\end{subequations}
where $\dif \mQ \coloneqq \prod_{a,b=1}^m \dif \re{[\mQ]_{ab}} \dif \img{[\mQ]_{ab}}$, the integral is taken over $\setC^{m\times m}$, and $\exp\set{- N\mai(\mQ)}$ indicates the density of ${\mas(\mQ)}$ which is written as
\begin{align}
\exp\set{-N\mai(\mQ)}=\sum_{\set{\bv_a}} \exp\set{ - N \Theta \set{\bv_a}} w(\mQ;\set{\bv_a}) \label{eq:26}
\end{align}
with the weight function $w(\mQ;\set{\bv_a})$ being
\begin{align}
w(\mQ;\set{\bv_a})=\prod_{a,b=1}^m &\delta( \re{N [\mQ]_{ab}-\bv_a^\her \bv_b}) \delta(\img{N [\mQ]_{ab} -\bv_a^\her \bv_b}). \label{eq:27}
\end{align}
We determine the density function in \eqref{eq:26} by replacing the impulse functions in \eqref{eq:27} with their inverse Laplace transform. In this case, after some lines of derivations, 
\begin{align}
\exp\set{-N\mai(\mQ)}= \int \exp\set{-N\tr{\mS\mQ} +N\mam(\mS)-\Delta_m} \dif \mS \label{eq:28}
\end{align}
where $\mS_{m\times m}$ is an square matrix enclosing the complex frequencies, 
\begin{align}
\dif \mS \coloneqq \prod_{a,b=1}^m (2\pi j)^{-2} \dif \re{[\mS]_{ab}} \dif \img{[\mS]_{ab}},
\end{align}
the integral is taken over $\setC^{m\times m}$, and ${\mam(\mS)}$ is defined as
\begin{align}
\mam(\mS) \coloneqq \frac{1}{N} \log \sum_{\set{\bv_a}} \exp\set{ \sum_{a,b=1}^m [\mS]_{ab} \bv_a^\her \bv_b - \beta \sum_{a=1}^m u(\bv_a) + h N \sum_{a=1}^m \sfM^{\setW}_f(\bv_a;N)}.
\end{align}
By partitioning the set $[N]$ as $[N]=\setW_N \cup \setW_N^{\sf C}$, where $\setW_N^{\sf C}$ indicates the complement of the index set $\setW_N$ \ac{wrt} $[N]$, ${\mam(\mS)}$ reduces to
\begin{align}
\mam(\mS) \coloneqq (1-\zeta)  &\log \sum_{\bvv} \exp\set{\bvv^\her \mS \bvv - \beta u(\bvv)} \nonumber \\  + \zeta &\log \sum_{\bvv} \exp\set{\bvv^\her \mS \bvv - \beta u(\bvv)+h \zeta^{-1} \sum\limits_{a=1}^m f(v_a)} 
\end{align}
with $\bvv\in\setX^m$ and $\zeta= \abs{\setW(N)}/N$. Here, one should note that $\bvv$ is an $m$-dimensional vector which is different from $\bv$. We refer to $\bvv$ as the vector of replicas. By substituting \eqref{eq:28} in \eqref{eq:25}, 
\begin{align}
f_{\rm M}(m)= \int \exp\set{-N \mg (\mT \mQ) -N \tr{\mS \mQ} +N \mam(\mS) - \Delta_m} \dif \mQ \dif\mS . \label{eq:29}
\end{align}
From \eqref{eq:29}, $\maf(\beta,h)$ is determined by substituting into \eqref{eq:18}. We further assume that the limits \ac{wrt} $N$ and $m$ commute which is a common assumption in replica analyses and is concluded when replica continuity holds. By substitution \eqref{eq:29} in \eqref{eq:18} and exchange of the limits, $\maf(\beta,h)$ reads 
\begin{align}
\hspace*{-1mm}\maf(\beta,h) = \lim_{m\downarrow 0} \lim_{N\uparrow \infty} \frac{1}{N} \frac{1}{m} \log \int \exp\set{-N \mg (\mT \mQ) -N \tr{\mS \mQ} +N \mam(\mS) - \Delta_m} \dif \mQ \dif\mS. \label{eq:repelkibudi}
\end{align}
To take the integral in \eqref{eq:repelkibudi}, we invoke Lemma~\ref{varahlemma} which leads us to conclude
\begin{align}
\maf(\beta,h) = - \lim_{m\downarrow 0} \frac{1}{m}  \left[ \mg (\mT \tilde{\mQ}) + \tr{\tilde{\mS} \tilde{\mQ}} - \mam(\tilde{\mS}) + \alpha \log (1+\rho \beta m) \right] \label{eq:repelkibudi2}
\end{align}
where we substitute $\Delta_m\coloneqq K \log (1+\rho \beta m)$. In \eqref{eq:repelkibudi2}, $(\tilde{\mS},\tilde{\mQ})$ is the saddle point of the exponent function $-\mg (\mT \mQ) - \tr{\mS \mQ} + \mam(\mS)$. We further simplify \eqref{eq:repelkibudi2} as
\begin{align}
\alpha \lim_{m\downarrow 0} \frac{1}{m}  \log \prant{ 1+\rho \beta m } &= \alpha \hspace*{.5mm} \frac{\partial}{\partial m} \log \prant{ 1+\rho \beta m }|_{m=0} = \alpha \rho \beta.
\end{align}
%
To derive the saddle point $(\tilde{\mS},\tilde{\mQ})$, we let the derivatives of the exponent function \ac{wrt} $\mQ$ and $\mS$ to zero. Using the standard matrix derivation, the derivative \ac{wrt} $\mQ$ results in 
\begin{align}
\tilde{\mS}=-\beta\mT \rmR_{\mJ} (-\beta \mT \tilde{\mQ}). \label{eq:ms}
\end{align}
Moreover, by taking the derivative \ac{wrt} $\mS$, and substituting $\tilde{\mS}$ as in \eqref{eq:ms}, we have
\begin{align}
\tilde{\mQ}&=\sum_{\bvv} \rmp^\beta(\bvv|\tilde{\mQ}) \bvv \bvv^\her \label{eq:33}
\end{align}
where we define the function $\rmp^\beta(\bvv|\tilde{\mQ})$ to be
\begin{align}
\rmp^\beta(\bvv|\tilde{\mQ})&=\dfrac{\exp\set{-\beta\left[\bvv^\her \mT \rmR_{\mJ} (-\beta \mT \tilde{\mQ}) \bvv+ u(\bvv)\right] }}{\sum_{\bvv} \exp\set{-\beta\left[\bvv^\her \mT \rmR_{\mJ} (-\beta \mT \tilde{\mQ}) \bvv+ u(\bvv)\right] }}. \label{eq:replica_distr}
\end{align}
and refer to it as the ``distribution of replicas''. By substituting $\maf (\beta, h)$ into \eqref{eq:10} and \eqref{eq:14}, the asymptotic marginal and distortion are determined as in Proposition~\ref{prop:general}.

\begin{proposition}[\bfseries General Replica Solution]
\label{prop:general}
Consider the nonlinear \ac{glse} precoder in Section \ref{sec:sys}, and define $\bvv_{m\times1}$ to be a random vector over $\setX^m$ with the distribution $\rmp_{\bvv}^\beta(\bvv;\mQ)$ given \eqref{eq:replica_distr}. 
%
Let $\tilde{\mQ}=\mQ^\star$ be a solution to the fixed-point equation in \eqref{eq:33}.
Then, under the replica continuity assumption, the asymptotic marginal of $f(\bx)$ is given by
\begin{align}
\sfM^{\setW}_f(\bx)=\lim_{\beta\uparrow\infty} \lim_{m\downarrow0} \sum_{\bvv} \rmp_{\bvv}^\beta(\bvv;\mQ^\star) \sfM^{\setT}_f(\bvv;m), \label{eq:marginkibudi}
\end{align}
and $\sfD (\rho) = \rho + {\alpha}^{-1} \lim\limits_{\beta\uparrow\infty}  \mad^\sfR(\beta)$ where $\mad^\sfR(\beta)$ is defined~as
\begin{align}
\hspace*{-2mm}\mad^\sfR(\beta) \coloneqq  &\frac{\partial}{\partial \beta} \left[ \lim_{m\downarrow 0} \frac{1}{m} \tr{\int_0^\beta \mT \mQ^\star \rmR_{\mJ}(-\omega \mT\mQ^\star) \dif\omega} \right] -\beta \lim_{m\downarrow 0} \frac{1}{m}  \tr{\mT \rmR_\mJ(-\beta\mT\mQ^\star) \frac{\partial\mQ^\star}{\partial \beta}} . \label{eq:dRkibudi}
\end{align}
\end{proposition}
Using Proposition~\ref{prop:general}, the asymptotics of a given \ac{glse} precoder are derived by determining $\mQ^\star$, and then, substituting it into \eqref{eq:marginkibudi} and \eqref{eq:dRkibudi}. The direct derivation of $\mQ^\star$, however, is not trivial. We therefore invoke the well-known approach in the literature in which the correlation matrix at the saddle-point is constructed step-wise by \ac{rs} and \ac{rsb} schemes.

\subsection{Solutions under Replica Symmetry and Replica Symmetry Breaking}
To derive the saddle-point matrix $\mQ^\star$, one should directly search over all possible matrices and find those which satisfy the fixed-point \eqref{eq:33}. This task is not feasible due to the following issues:
\begin{inparaenum}
\item the search is computationally complex, and
\item the derived solutions to the fixed-point equations are not guaranteed to result in an analytic expression for $f_{\rm M}(m)$.
\end{inparaenum}
These issues together are addressed in the literature of spin glasses by invoking a well-known trick. A trick is to restrict the set of matrices, over which we search for the saddle-point, to be of a given structure. The structure is parameterized with several parameters which are to be found such that the matrix with assumed structure satisfies the fixed-point equation. To clarify the idea further, assume that we suppose $\tilde{\mQ}=\mA(q_1, \ldots, q_L)$ in which $\mA(q_1, \ldots, q_L)$ is an $m\times m$ matrix which is parameterized by $q_1, \ldots, q_L$. Here, $L$ is a constant number which does not depend on $m$. In this case, by inserting $\mA(q_1, \ldots, q_L)$ in \eqref{eq:33}, the fixed-point equation reduces to $L$ coupled equations which determine $q_1^\star, \ldots, q_L^\star$, and consequently, ${\mQ}^\star=\mA(q_1^\star, \ldots, q_L^\star)$. Using this trick, the problem of finding the saddle-point can be feasibly followed in an analytic way \cite{mezard2009information}. There are however two main questions in this respect:
\begin{inparaenum}
\item Which structure on $\mQ^\star$ should be considered?
\item How can we assure that a correlation matrix at the saddle-point has the given structure?
\end{inparaenum}
The former question has been answered in the literature of spin glasses. In fact, due to some specific properties of the exponent function the initial conjecture on the saddle-point structure is \ac{rs} which indicates that ${\mQ}^\star$ reads 
\begin{align}
{\mQ^\star}= \frac{\chi}{\beta} \mI_m + \sfp \mone_m \label{eq:rs}
\end{align}
for some $\chi$ and $\sfp$. It is however not guaranteed that the saddle-point matrix has the structure of this form. This latter statement brings the need to answer the second question. To check if the saddle-point has the \ac{rs} structure, a series of stability analyses have been considered in the literature \cite{mezard2009information,dotsenko2005introduction,nishimori2001statistical}. The analyses indicate that, depending on the Hamiltonian, the \ac{rs} structure may not address the exact saddle point and in fact gives a lower bound on that. To find the saddle-point in these cases, the assumed structure on ${\mQ^\star}$ has to be modified. Parisi proposed the \ac{rsb} scheme in \cite{parisi1980sequence} which starts from \ac{rs} and widens the structure recursively\footnote{In the context of the replica method, this recursive widening is called ``symmetry breaking''.}. Using Parisi's scheme \cite{parisi1980sequence}, after one step of recursion, one gets 
\begin{align}
\mQ^\star= \frac{\chi}{\beta} \mI_m + \sfc \hspace*{.5mm} \mI_{\frac{m\beta}{\mu}} \otimes \mone_{\frac{\mu}{\beta}} +  \sfp \mone_m \label{eq:1rsb}
\end{align}
Here, $\mQ^\star$ is controlled by $\chi$, $\sfp$, $\sfc$ and $\mu$. More steps of recursion result in wider classes of structures. The solution obtained by $b$ recursions of \ac{rsb} is referred to as the $b$-steps \ac{rsb} solution and when $b\uparrow\infty$ is called the full-\ac{rsb} solution. In general, it is shown that \ac{rs}, as well as $b$-steps \ac{rsb}, give a lower bound on the exact solution which gets tighter as $b$ grows. In the sequel, we determine the \ac{rs} as well as one-step \ac{rsb} solution which conclude Propositions~\ref{thm:1} and \ref{thm:2}. The extension to to more steps of \ac{rsb} is also discussed in Appendix~\ref{app:b-rsb}. For derivation, we invoke the systematic approach illustrated in \cite[Appendices B-D]{bereyhi2016statistical}.


\subsubsection*{Replica Symmetric Solution}
By substituting \eqref{eq:rs} in Proposition~\ref{prop:general}, we have
\begin{align}
\rmp^\beta(\bvv|\tilde{\mQ}) = \int \prod\limits_{a=1}^m \frac{ \exp\set{-\beta\mae_\rs(\vv_a|s_0)} }{\int \left[ \sum_{\vv} \exp\set{-\beta\mae_\rs(\vv|s_0)} \right]^m \md s_0} \md s_0, \label{eq:p_rs}
\end{align}
where $\md s_0 \coloneqq {\exp\set{-\abs{s_0}^2}} \dfrac{\dif s_0}{\pi}$, and the function $\mae_\rs(\cdot|s_0):\setX\mapsto\setR^+$ reads
\begin{align}
\mae_\rs(\vv|s_0)\coloneqq \frac{1}{\xi} \abs{\vv}^2 - 2 \varsigma \re{ \vv s_0^*} + u(\vv) \label{eq:e_rs}
\end{align}
with $\xi=\left[\rmR_\mJ(-\chi)\right]^{-1}$, and $\varsigma$ being
\begin{align}
\varsigma^2= \frac{\partial}{\partial \chi} \left[ ( \rho \chi- \sfp ) \rmR_\mJ(-\chi) \right] .
\end{align}
By replacing the \ac{rs} structure into the fixed-point equation \eqref{eq:33} and taking limits $\beta\uparrow\infty$ and $m\downarrow 0$, $\chi$ and $\sfp$ at the saddle-point are found to be
\begin{subequations}
\begin{align}
\sfp&=\int \abs{\arg\min_{\vv} \mae_\rs(\vv|s_0)}^2 \hspace*{.5mm} \md s_0, \label{eq:46a} \\
\chi&= \frac{1}{\varsigma}\int \re{\arg\min_{\vv} \mae_\rs(\vv|s_0) \hspace*{.8mm} s_0^*} \hspace*{.5mm} \md s_0. \label{eq:46b}
\end{align}
\end{subequations}
Finally, by determining the asymptotic distortion for the correlation matrix of the form \eqref{eq:rs} and defining $\rho^\rs= \xi^2 \varsigma^2$, Proposition \ref{thm:1} is concluded. The detailed lines of derivations for \eqref{eq:p_rs}-\eqref{eq:46b} take the same steps as in \cite[Appendix B]{bereyhi2016statistical} and are skipped due to similarity.

\subsubsection*{One-step Replica Symmetry Breaking Solution}
Considering \ac{rsb} scheme with one step of recursion, the distribution of replicas is given by
\begin{align}
\rmp^\beta(\bvv|\tilde{\mQ}) = \int \prod\limits_{j=1}^{\frac{m\beta}{\mu}} \int \prod_{a\in\setS_j} \frac{ \exp\set{-\beta\mae_\rsb(\vv_a|s_0, s_1)} }{\int \left[ \int \left[ \sum_{\vv} \exp\set{-\beta\mae_\rsb (\vv|s_0,s_1)} \right]^{\frac{\mu}{\beta}} \md s_1 \right]^{\frac{m \beta}{\mu}} \md s_0 } \md s_0 \md s_1, \label{eq:p_rsb}
\end{align}
where $\setS_j= [\frac{(j-1)\mu}{\beta}+1 : \frac{j \mu}{\beta}]$ and 
\begin{align}
\mae_\rsb(\vv|s_0,s_1) \coloneqq \frac{1}{\xi} \abs{\vv}^2 - 2\varsigma_0 \re{ \vv s_0^*} + 2 \varsigma_1 \re{\vv s_1^*} + u(\vv) \label{eq:e_rsb}
\end{align}
with $\xi=\left[\rmR_\mJ(-\chi)\right]^{-1}$ and $\varsigma_0$ and $\varsigma_1$ being
\begin{subequations}
\begin{align}
\varsigma_0^2 &= \frac{\partial}{\partial \chi} \left[ ( \rho \chi + \rho \mu\sfc - \sfp ) \rmR_\mJ(-\chi-\mu\sfc) \right] \\
\varsigma_1^2 &= \frac{1}{\mu} \left[ \rmR_\mJ(-\chi)-\rmR_\mJ(-\chi-\mu\sfc) \right].
\end{align}
\end{subequations}
By substituting into \eqref{eq:33} and taking the limits $m\downarrow 0$ and $\beta \uparrow \infty$, the control parameters $\chi$, $\sfc$, $\sfp$ and $\mu$ at the saddle-point are found to satisfy the fixed-point equations
\begin{subequations}
\begin{align}
\sfp + \sfc &= \int \int \abs{\arg\min_{\vv} \mae_\rsb(\vv|s_0,s_1)}^2 \hspace*{.5mm} \tilde{\Lambda}\prant{s_0,s_1} \md s_0 \md s_1, \label{eq:sad-rsb1} \\
\chi+ \mu \sfc &= \frac{1}{\varsigma_0}\int \int \re{\arg\min_{\vv} \mae_\rsb (\vv|s_0,s_1) \hspace*{.8mm} s_0^*} \hspace*{.5mm} \tilde{\Lambda}\prant{s_0,s_1} \md s_0 \md s_1, \label{eq:sad-rsb2}\\
\chi+ \mu \sfc + \mu \sfp &= \frac{1}{\varsigma_1}\int \int \re{\arg\min_{\vv} \mae_\rsb (\vv|s_0,s_1) \hspace*{.8mm} s_1^*} \hspace*{.5mm} \tilde{\Lambda}\prant{s_0,s_1} \md s_0 \md s_1, \label{eq:sad-rsb3}
\end{align}
and
\begin{align}
\mu^2 \sfp \varsigma_1 + \frac{\mu \sfc}{\xi} - \int_\chi^{\chi+\mu \sfc} \rmR_{\mJ}(-\omega) \dif \omega = \int \log \int {\Lambda}\prant{s_0,s_1} \md s_1 \md s_0
\end{align}
\end{subequations}
where $\Lambda \prant{s_0,s_1}$ is defined as
\begin{align}
\Lambda \prant{s_0,s_1} \coloneqq \exp\set{-\mu \min_{\vv} \mae_\rsb (\vv|s_0,s_1)},
\end{align}
and  $\tilde{\Lambda} \prant{s_0,s_1}$ reads
\begin{align}
\tilde{\Lambda} \prant{s_0,s_1} \coloneqq \frac{\Lambda \prant{s_0,s_1}}{\int \Lambda \prant{s_0,s_1} \md s_1}.
\end{align}
By defining the variances $\rho^\rs= \xi^2 \varsigma_0^2$ and $\rho_1^\rsb= \xi^2 \varsigma_1^2$ and the conditional distribution
\begin{align}
\rmp(s_1|s_0)= \tilde{\Lambda} \prant{s_0,s_1}  \phi(s_1), \label{eq:sad-rsb-end}
\end{align}
Proposition~\ref{thm:2} is concluded. Detailed derivations are similar to those given in \cite[Appendix C]{bereyhi2016statistical}.

\section{Conclusion}
We have proposed \ac{glse} precoding for the downlink of massive \ac{mimo} systems. This precoding scheme addresses several instantaneous constraints on the transmit signals jointly and outperforms conventional approaches. 
Using the optimal \ac{glse} precoder for joint antenna selection and power control at the transmitter, the number of active transmit antennas can be reduced up to $54\%$ compared to classical algorithms. This enhancement reduces to $41\%$ when a computationally efficient \ac{glse} precoder with $\ell_1$-norm penalty is employed. \ac{glse} precoding further lets us construct transmit signals from discrete constellations, such as $M$-PSK, while imposing side constraints, e.g., limited number of active antennas, on the signal.

\ac{glse} precoding opens several directions for future studies. Design and analysis of low complexity algorithms for implementation of \ac{glse} precoders is one possible direction. For this aim, approximate message passing algorithms can be employed to address iteratively \ac{glse} precoding. This study has been already started, and some initial results have been demonstrated in \cite{bereyhi2018precoding}. Another direction for future work is to extend this precoding scheme to precode~simultaneously a block of parallel data streams. By this extension, further hardware limitations in some practical scenarios such as OFDMA transmission can be addressed. The work in this direction is currently ongoing.


\appendices
\newpage
\section{Extension to $b$-steps \ac{rsb}}
\label{app:b-rsb}
After $b$ steps of recursion, the structure given by the \ac{rsb} scheme can be written as
\begin{align}
\tilde{\mQ}= \frac{\chi}{\beta} \mI_m + \sum_{\kappa=1}^b \sfc_\kappa\hspace*{.5mm} \mI_{\frac{m\beta}{\mu_\kappa}} \otimes \mone_{\frac{\mu_\kappa}{\beta}} +  \sfp \mone_m \label{eq:rsb}
\end{align}
In this case, $\tilde{\mQ}$ is controlled by $\chi$ and $\sfp$ and the sequences $\set{\sfc_\kappa}$ and $\set{\mu_\kappa}$ for $\kappa\in[1:b]$. As $b$ grows large \eqref{eq:rsb} cover all possible saddle-points, and thus, the saddle-point is derived precisely. The solution under $b\uparrow\infty$ is called the \textit{full-\ac{rsb}} solution, and for some given setups has been derived, e.g., Sherrington-Kirkpatrick model \cite{mezard1987spin}. To derive the $b$-steps \ac{rsb} solution, one needs to substitute $\tilde{\mQ}$ in Proposition~\eqref{prop:general} and take the limits \ac{wrt} $\beta$ and $m$. The derivations in this case are of more complicated form. By taking the same steps as in \cite[Appendix D]{bereyhi2016statistical}, one concludes that the solution is of an extended form of the one-step \ac{rsb} solution with 
\begin{align}
s^\dec=s^\rs+\sum_{\kappa=1}^b s_\kappa^\rsb
\end{align}
in which $s^\rs\sim\mathcal{CN}\prant{0,\rho^\rs}$ and $s_\kappa^\rsb$ is distributed conditioned to $s^\rs$ and $\set{s_1^\rsb,\ldots,s_{\kappa-1}^\rsb}$. The fixed-point equations are moreover of the extended form of the one-step \ac{rsb} case. More detailed illustrations can be found in \cite{bereyhi2016statistical}.

\section{Asymptotic Bound on the Distortion of $M$-PSK Transmission with \ac{tas}}
\label{app:lower}

Consider an \ac{iid} flat Rayleigh fading channel and assume that the entries of the transmit vector $\bx$ are either zero or $M$-PSK. Let the fraction of non-zero entries be $\eta$ and denote their amplitude by $\sqrt{P}$. Define the random variable $D(\bx)\coloneqq \norm{\mH \bx - \sqrt{\rho} \bs}^2/K$ and the probability $p_{\min}$ to be 
\begin{align}
p_{\min}=\Pr \set{ \min_{\bx} D(\bx) \leq \sfD_\ell }
\end{align}
In this case, one can invoke the union bound and write
\begin{align}
p_{\min}&=\Pr \set{ \cup_{\bx} D(\bx) \leq \sfD_\ell } \leq \sum_{\bx} \Pr \set{  D(\bx) \leq \sfD_\ell }. \label{eq:LOWE_1}
\end{align}
To determine the upper bound in \eqref{eq:LOWE_1}, we determine the distribution of $D(\bx)$ in the large system limit. For a given vector $\bx$, the entry $[\mH\bx]_k$ for $k\in\left[1:K\right]$ is the sum of $\eta N$ independent complex Gaussian random variables via complex coefficients whose~amplitude~is~$\sqrt{P}$. Noting that $\mH$ has \ac{iid} entries with zero-mean and variance $1/N$, $[\mH\bx]_k$ is concluded to be normally distributed with zero mean and variance $\eta P$. Consequently, each entry of $\mH \bx - \sqrt{\rho} \bs$ is a complex Gaussian random variable with zero-mean and variance $ \rho+\eta P$, and thus, $D(\bx)$ is distributed with
\begin{align}
f(d)=\frac{K^K d^{K-1}}{(\rho+\eta P)^K (K-1)!} \exp\set{ - \frac{K d}{\rho+\eta P} }
\end{align}
for all $\bx$ in the corresponding constellation $\setX^N$. As the result, \eqref{eq:LOWE_1} reads
\begin{subequations}
\begin{align}
p_{\min} &\leq \abs{\setX}^N \int_{0}^{\sfD_\ell} f(d) \dif d \\
&=(1+M)^N \int_{0}^{\sfD_\ell} f(d) \dif d.
\end{align}
\end{subequations}
Noting that $f(d)$ is an increasing function within the vicinity of zero, one can further write
\begin{align}
p_{\min} &\leq (1+M)^N \frac{K^K \sfD_\ell^{K-1} e^{K-1}}{\sqrt{2\pi}(\rho+\eta P)^K (K-1)^{K-1/2} } \exp\set{ - \frac{K {\sfD_\ell}}{\rho+\eta P} }
\end{align}
where we have replaced $(K-1)!$ with its lower bound
\begin{align}
(K-1)! \geq \sqrt{2\pi} (K-1)^{K-1/2} e^{-K+1}
\end{align}
given in \cite{robbins1955remark}. As the logarithm function is an increasing function one can write
\begin{align}
\log p_{\min} &\leq K\left[ 1- \frac{\sfD_\ell}{\rho+\eta P} + \frac{1}{\alpha} \log (1+M) + \log \frac{\sfD_\ell}{\rho+\eta P} \right] + \epsilon_K
\end{align}
where $\epsilon_K \downarrow 0$ as $K\uparrow \infty$. To make sure that the distortion in the large-system limit is bounded from below by $\sfD_\ell$, we need the probability $p_{\min}$ tend to zero in the large limit which holds when
\begin{align}
1- \frac{\sfD_\ell}{\rho+\eta P} + \frac{1}{\alpha} \log (1+M) + \log \frac{\sfD_\ell}{\rho+\eta P} < 0. \label{eq:sfD_LOWER_neu}
\end{align}
Therefore, for any $\sfD_\ell$ which satisfy \eqref{eq:sfD_LOWER_neu} the asymptotic distortion is greater than $\sfD_\ell$. Consequently, the lower bound in Lemma~\ref{lemma2} if found by choosing the maximum $\sfD_\ell$ fulfilling \eqref{eq:sfD_LOWER_neu}. This is given by replacing $<$ with equality.

\section{Constant Envelope Transmission with \ac{tas}}
\label{app:entropy}
To investigate the constant envelope transmission with \ac{tas}, one can start from the case with $M$-PSK constellation and take the limit $M\uparrow\infty$. Considering the derivations in Section~\ref{sec:MPSK}, one can show that in the large-system limit of $M$ 
\begin{align}
\frac{2k^\star \pi}{M} \to \sphericalangle s^\rs,
\end{align}
and therefore, the $\Theta\prant{k^\star | \sphericalangle s^\rs} = 1$ which reduces the threshold to
\begin{align}
\tau_{\sfd} = \frac{\sqrt{P}}{2} \prant{1+\xi\lambda}.
\end{align}
In this case, the asymptotic fraction of the active antennas reads
\begin{align}
\eta = \exp\set{ - \frac{\tau_{\sfd}^2}{\rho^\rs }}
\end{align}
and $\sfp=P\eta$. Moreover, the fixed-point equation reduces to
\begin{align}
\frac{\alpha \chi \rho^\rs}{ 1+\chi} =  \sqrt{P} \tau_{\sfd} \eta +  \sqrt{\pi \rho^\rs P} \hspace*{1mm} \rmQ\prant{ \sqrt{\frac{2}{\rho^\rs}} \tau_{\sfd} }.
\end{align}
Finally, the asymptotic distortion is determined by \eqref{eq:sfD0}. Here, one should know that in this case the transmit \ac{papr} is not zero, since the transmit vector has $\eta N$ number of zero-elements. The non-zero transmit symbols however have no amplitude variation.

\section{Block-wise \ac{glse} Precoding}
\label{app:block_GLSE}
The basic form of \ac{glse} precoding constructs the transmit signal \textit{slot-wise}. This means that for each sub-channel in the frequency domain, the transmit signal in each transmission time slot is constructed by running the algorithm completely. Such an approach needs to be extended to \textit{block-wise} algorithms for practical systems due to two major issues:
\begin{enumerate}
\item The slot-wise precoding poses a large computational load into the system. In addition to this, several applications of \ac{glse} precoding, impose further complexities into the system which cannot be updated within each slot. An example is the \ac{glse} precoder which addresses the antenna selection task. In this case, slot-wise precoding needs the \ac{rf}-chains to be switched at the beginning of each transmission time slot. For high data rates, such a task is neither feasible nor efficient, since the switching speed is limited for a given insertion loss and high-speed switching cause large loss into the system. As the result, the precoding algorithm needs to be updated for a larger period of time, e.g., within each coherence~time~interval.
\item In practice, \ac{mimo} systems employ multi-carrier modulation schemes, e.g., \ac{ofdm}, for transmission over channels with frequency selective fading. In these schemes, the data stream is divided into parallel sub-streams. These sub-streams need to be precoded for different sub-channels which correspond to the various sub-carriers. For these systems, using slot-wise precoding means to allocate a separate precoding module to each sub-carrier which leads to a large computational burden into the transmitter.
\end{enumerate}
In order to address both of these issues, \ac{glse} precoding needs to be extended to a block of data streams. To formulate this generalization formally, consider a set of independent data vectors $\set{\bs_1, \ldots, \bs_B }$ which represent the data streams over distinct $B$ dimensions, e.g., data streams over distinct transmission time slots or various sub-carriers. Let $\mH_b\setC^{K\times N}$, for $b\in[B]$, represent the channel matrix corresponding to dimension $b$:
\begin{enumerate}
\item For example, when $\set{\bs_1, \ldots, \bs_B }$ represent data streams over multiple transmission time slots in a single coherence time interval, $\mH_b=\mH$, for $b\in[B]$, where $\mH$ denotes the channel matrix in the corresponding coherence time interval.
\item Assuming that $\set{\bs_1, \ldots, \bs_B }$ denote parallel data streams corresponding to different sub-carriers in a multi-carrier modulation scheme, $\mH_b$ describes the effective channel matrix between the transmitter and users in the sub-channel corresponding to the~$b$-th~sub-carrier.
\end{enumerate}
In this case, block-wise \ac{glse} precoding with penalty $u(\cdot):\setX^N \mapsto \setR$ and support $\setX$ reads
\begin{align}
\glse_{\rm block}\prant{\set{\bs_b}_{b=1}^B |\rho,\set{\mH_b}_{b=1}^B}=\argmin_{\bv_1, \ldots, \bv_B \in {\setX^N}} \sum_{b=1}^B \norm{\mH_b \bv_b-\sqrt{\rho} \hspace*{.3mm} \bs_b}^2 + u_{\rm block}(\bv_1,\ldots,\bv_B) \label{eq:block_GLSE_B}
\end{align}
for some power control factors $\rho$ and some block penalty $u_{\rm block}(\cdot)$. Here, the precoder returns a block of transmit vectors $\bx_1, \ldots,\bx_B$. 

The block-wise \ac{glse} precoder simultaneously maps a block of data streams to a set of transmit signals, such that the total distortion is minimized given some desired constraints. As the result, one can employ this extended scheme in practical scenarios. It is straightforward to show that this block-wise class of precoders can be formulated as basic \ac{glse} precoding when the penalty function decouples over the blocks. For this aim, assume that 
\begin{align}
u_{\rm block} (\bv_1,\ldots,\bv_B) = \sum_{b=1}^B u(\bv_b),
\end{align}
for some penalty $u(\cdot)$. Define the vectors $\bs_{\rm t} \in \setC^{KB\times 1}$ and $\bv_{\rm t} \in \setX^{NB\times 1}$ as
\begin{subequations}
\begin{align}
\bs_{\rm t} &= [\bs_1^\trp, \ldots, \bs_B^\trp]^\trp,\\
\bv_{\rm t} &= [\bv_1^\trp, \ldots, \bv_B^\trp]^\trp,
\end{align}
\end{subequations}
and the block-diagonal matrix $\mH_{\rm t}=\mathrm{diag}\left( \mH_1, \ldots,\mH_B\right)\in \setC^{KB\times NB}$. The block-wise precoding scheme in \eqref{eq:block_GLSE_B} in this case reads
\begin{subequations}
\begin{align}
\glse_{\rm block}\prant{\set{\bs_b}_{b=1}^B |\rho,\set{\mH_b}_{b=1}^B}&=\argmin_{\bv_t \in {\setX^NB}} \norm{ \mH_{\rm t} \bv_{\rm t}-\sqrt{\rho} \hspace*{.3mm} \bs_{\rm t} }^2 + u(\bv_{\rm t} )\\
&= \glse\prant{\bs_{\rm t}  |\rho, \mH_{\rm t} }. \label{eq:final_EQQQ}
\end{align}
\end{subequations}
As the result, the results for the basic setups is generalized by replacing the channel matrix and the data vector with $\mH_{\rm t}$ and $\bs_{\rm t}$. For some applications, the decoupling over the block is not possible. In this case, the approaches from the literature of distributed sensing networks\footnote{See for example \cite{bereyhi2018theoretical} and the references therein.} can be employed to investigate the impacts of grouping on the precoding performance. Currently, the work in this direction is under the way.


\section{A Benchmark \ac{tas} Algorithm}
\label{app:tas_alg}
Assume that $L$ antennas are to be selected out of the total $N$ antennas available at the transmitter. A conventional low-complexity approach for \ac{tas} is to select the transmit antennas which have the strongest channel gains: Let $\mh_n\in \setC^{K\times 1}$ denote the $n$-th column of $\mH$ which corresponds to transmit antenna $n$. We define $\set{w_1, \ldots, w_N}$ to be a permutation of $[N]$ for which
\begin{align}
\norm{\mh_{w_1}} \geq \ldots \geq \norm{\mh_{w_N}}.
\end{align}
In this case, the \ac{tas} algorithm selects the $L$ transmit antennas which correspond to $\set{w_1, \ldots, w_L}$.


\bibliography{ref}
\bibliographystyle{IEEEtran}

\end{document}